\documentclass[twocolumn,showpacs,preprintnumbers,prd,superscriptaddress,nofootinbib,floatfix]{revtex4}
\usepackage{amsmath}
\usepackage{amsfonts}
\usepackage{graphicx}
\usepackage{hyperref}

\begin{document}

\title{The dark degeneracy and  interacting cosmic components}
\author{Alejandro Aviles}
\email{aviles@ciencias.unam.mx}
\affiliation{Instituto de Ciencias Nucleares, UNAM, M\'exico}
\affiliation{Depto. de F\'isica, Instituto Nacional de Investigaciones Nucleares, M\'exico}
\author{Jorge L. Cervantes-Cota}
\email{jorge.cervantes@inin.gob.mx}
\affiliation{Depto. de F\'isica, Instituto Nacional de Investigaciones Nucleares, M\'exico}

\hyphenation{quin-tes-sen-ce}
\hyphenation{His-to-ry}

\begin{abstract}
We study some properties of the dark degeneracy, which is the fact that what we measure in gravitational experiments is the energy momentum tensor of 
the total dark sector, and any split into components (as in dark matter and dark energy) is arbitrary. 
In fact, just one dark fluid is necessary to obtain exactly the same cosmological and astrophysical phenomenology
as the $\Lambda$CDM model. We work explicitly the first-order perturbation theory and show that beyond the linear order the dark degeneracy 
is preserved under some general assumptions. Then, we construct the dark fluid from a collection of
interacting fluids. Finally, we  try to break the degeneracy with a general class of couplings to baryonic matter. Nonetheless, 
we show that these interactions can also be understood in the context of the $\Lambda$CDM model as between dark matter and baryons.
For this last investigation we choose two independent parametrizations for the interactions, one inspired by electromagnetism and 
the other by chameleon theories. Then, we constrain them with a joint analysis of CMB and supernovae observational data.
\end{abstract}
\date{October 12, 2011}
\maketitle

\section{Introduction}

Current cosmological observations indicate that about 96$\%$ of the total energy content of our Universe is made of yet unknown dark components and
only 4$\%$ is made of particles of the standard model. Among them, there are precision measurements of anisotropies in the Cosmic Microwave Background (CMB) 
radiation \cite{WMAPmission,Komatsu,Larson}, baryon acoustic oscillations \cite{Eisenstein05,Percival07}, and Type Ia Supernovae \cite{Riess98,Perlmutter99,Union2}. For a recent
review of the nowadays status of cosmology see, e.g., \cite{JorgeSmoot}.

Usually this {\it dark fluid} is separated into two components: a clustering {\it dark matter} piece
responsible for forming structure in the Universe and a nonclustering {\it dark energy} with a negative pressure that is responsible for the current accelerated cosmic
expansion. 

Moreover, in the concordance model of cosmology --the so-called Lambda Cold Dark Matter ($\Lambda$CDM) model-- the dark energy 
is of  geometric nature and enters as a constant at the Lagrangian level of Einstein's gravitational theory. For gravity matters, the 
cosmological constant is indistinguishable from the vacuum contribution of quantum fields, and here appears an intriguing problem, the sum of both contributions 
is about 1 over $10^{120}$ times the latter \cite{Weinberg}. A second conceptual problem with the $\Lambda$CDM model comes out when one considers the ratio of 
dark energy over dark matter energy densities, which in spite of growing with the third power of the scale factor of the Universe,  its value today is of  order one. This is
the so-called {\it coincidence problem}.
As a consequence, a plethora of alternative proposals has appeared in the literature; 
as a short sample see \cite{RatraPeeb88,Caldwell98,Copeland98,Kessence,Carrol03,Sotiriou08,DGP,CopSamTsu06,Tsujikawa2010,Saridakis2010}.

In this paper we argue that the separation of the dark fluid into dark matter and dark energy  is arbitrary and is only favored for historical reasons and computational 
simplicity. After all, we define the dark fluid as our lack of knowledge as\footnote{We use natural units, $\hbar =1$ and $c=1$, otherwise is stated.}
\begin{equation}
 T_{\mu\nu}^{dark} = \frac{1}{8 \pi G } G_{\mu\nu} -   T_{\mu\nu}^{obs},
\end{equation}
where $G_{\mu\nu}\,$ comes from the observed geometry of the Universe and  $T_{\mu\nu}^{obs}\,$ from its observed energy content.
In fact, the dark sector could be composed by a large zoo of particles with complicated interactions between them. Or, it could be even
just one exotic unknown  dark fluid.
This property has been called {\it dark degeneracy} by M. Kunz \cite{Kunz09}; see also \cite{Hu99,Rubano02,Wasserman02,Liddle2006,Kunz20092,Aviles,Reyes2011}.

In this paper we assume an astrophysical perspective  to the dark fluid by defining it as a barotropic fluid 
with  speed of sound equal to zero, as in \cite{Beca2007,Balbi07} and more recently in \cite{Quevedo}; for  similar approaches see \cite{Linder09,Arbey2005,Avelino03}. 

Making the speed of sound equal to zero ensures that dark fluid energy density perturbations will grow at all scales, but at the same time, it allows 
the fluid to have a nonzero pressure. This is  quite in contrast with canonical 
scalar fields as quintessence, for which the speed of its 
perturbations is equal to one, a feature that makes  dark energy perturbations to be  quickly damped, so they do not grow inside the horizon. 

Interactions within the dark sector has been studied largely in the literature 
\cite{Amendola2000,Mangano2003,FarPeeb04,Valviviita2008,Bean08,Caldera09,Costa09,Gavela2010,He2009}, mainly as a mechanism to solve the  coincidence problem. 
On the other hand, interactions between dark matter and the standard model of particles are expected, and have been studied also on several occasions. In fact, 
the weakly interacting massive particles (WIMPs) paradigm has emerged as the predominant scenario to solve the missing mass problem 
\cite{Goldberg1983,Jungman1996,Bertone2005,Yuksel2007,Beltran2009}.
Nevertheless, other alternatives have been proposed. Among them, special attention has been attracted by the strong interacting dark matter scenarios 
\cite{Spergel2000,Wandelt2001,Cyburt2002,Chen2002,Erickcek2007,Mack2007}, in which
dark matter interacts with itself and with baryons \cite{Wandelt2001} through the strong force. It has been shown \cite{Spergel2000,Dave2001} that this alternative 
can alleviate the cuspy halos and overdensity of substructure problems that appear in N-body simulations based on the standard $\Lambda$CDM model \cite{Navarro}.

In this paper we develop a general class of couplings of the dark fluid to the standard model of particles, and show that they can be also understood as interactions of  
dark matter to baryons. Then, we impose constraints by using CMB anisotropies and supernovae observations.
 
This work is organized as follows: In Sec. II we define the dark fluid from the properties of the dark matter itself. In Sec. III we work out the
cosmological background solutions for the dark fluid and show that they are identical to the $\Lambda$CDM model ones, leading to the dark degeneracy.
In Sec. IV we show that the dark degeneracy is preserved under some assumptions when one goes beyond  zero order in cosmological perturbation theory. 
In Sec. V we work a multifluid description of the dark sector, allowing interaction among its components. In Sec. VI we extend these interactions to baryons
to try to break the dark degeneracy. Finally, in Sec. VII, we present our conclusions.

\section{The sound speed of the dark fluid}

Considering Newtonian gravity for the moment, the overdensities of matter in an expanding Universe follow the equation in Fourier space

\begin{equation}
 \delta''  + 2 H \delta' + \left(c^2_s k^2_{\text{phys}} - 4 \pi G \rho \right) \delta = 0,
\end{equation}
where $k_{\text{phys}}$ is a physical wavelength (contrary to the comoving wavelength which we will use in the following sections) related
to a physical length scale by $k_{\text{phys}} = 2 \pi/ l_{\text{phys}}$, $c^2_s$ is the speed of sound of the fluid under consideration and 
prime means derivative with respect to cosmic time.
In astrophysical scales it is safe to set $H = 0$ and then, the scale
factor is equal to a constant that we make equal to one for this discussion matters. It follows that there  is a threshold scale, the Jeans length, given by
\begin{equation}
 l_J = c_s  \sqrt{\frac{\pi}{G \rho}}
\end{equation}
for which  perturbations with physical length above it, $l_{\text{phys}} > l_{J}$, grow by gravitational collapse, and perturbations with $l_{\text{phys}}<l_J$ 
develop acoustic oscillations.

We expect to have dark matter structure at a wide range of scales: from the largest cosmological structure to galaxies. In fact even dwarf galaxies need to 
have dark matter halos 
in order to stabilize their disks and to account for the necessary gravity sources to obtain the observed flat rotation curves which follow the luminous
matter in there. The only way to guarantee that dark matter perturbations grow at all
scales is demanding that the Jeans length be equal to zero.

For barotropic fluids the adiabatic speed of sound coincides with the speed at which perturbations propagate in the fluid.
We define the dark fluid as a barotropic fluid with an adiabatic speed of sound equal to zero,

\begin{equation}
 c^2_s=0,   \label{dfdef}
\end{equation}
and that, at a first approximation, does not interact with particles of the standard model.\footnote{The barotropic condition dismisses, for instance, 
the possibility of scalar fields for which although the speed of propagation of its perturbations is equal to the speed of light, 
this does not coincide with its adiabatic speed of sound, which indeed could be zero.} Without lost of generality we can write its equation of state as

\begin{equation}
 P(\rho) = w(\rho)\rho
\end{equation}
where $w$ is its equation of state parameter and is a function of the energy density $\rho$  only. 
Thus we have $c^2_s = (\partial P/\partial \rho)_s = dP/d\rho$.
Using Eq. (\ref{dfdef}) we obtain that the equation of state parameter is solved by the equation $\rho dw/d \rho + w = 0$, whose solution is
\begin{equation}
 w = -\frac{\mathcal{C}}{\rho}, \label{EOSdf}
\end{equation}
with $\mathcal{C}$ a constant and the minus sign is set for later convenience.  This means that the pressure is a constant
\begin{equation}
 P = P_0 = -\mathcal{C}.
\end{equation}
Astrophysical observations constrain this value to be very small, $|P| \ll \rho_A$, where $\rho_A$ is the energy density of typical astrophysical scales where
dark matter has been detected. Usually, it is
assumed that dark matter is pressureless, but this is by not means necessary, for instance it 
could be the case that $|P| \sim \rho_{c0}$,  where  $\rho_{c0}$ is a typical cosmological  energy density scale at present, without getting in contradiction with observations. 
In fact, this is the entrance that leads us to consider the dark fluid to be dark energy as well as dark matter.

\section{Dark fluid as dark energy}

Now, let us consider a Friedmann-Robertson-Walker description of the Universe at very large scales, filled with standard model particles ($b$, $\gamma$, ...) 
and with the above-defined  dark fluid, from now on labeled by $d$. The evolution equations of such a
Universe are

\begin{equation}
 H^2 = \frac{8 \pi G}{3}(\rho_d + \rho_b + \rho_{\gamma}), \label{s1}
\end{equation}
\begin{equation}
 \rho_b' + 3H\rho_b =0, \label{s2}
\end{equation}
\begin{equation}
\rho_{\gamma}' + 4 H \rho_{\gamma} =0, \label{s3}
\end{equation}
and
\begin{equation}
\rho_d' + 3H(1+w_d)  \rho_d  = 0, \label{s4}
\end{equation}
where $H\equiv a'/a$ is the Hubble factor. Equations (\ref{s2}) and (\ref{s3}) give 
$\rho_b = \rho_{b0} a^{-3}$ and $\rho_{\gamma} = \rho_{\gamma 0} a^{-4}$, where a subindex $0$ means that the quantity under consideration
is evaluated at present time, and we have normalized the scale factor to be equal to one today, $a_0 = 1$. Integration of Eq. (\ref{s4}) gives,
using Eq. (\ref{EOSdf}),

\begin{equation}
\rho_d = \frac{\rho_{d0}}{1+\mathcal{K}}\left( 1+ \frac{\mathcal{K}}{a^3} \right),
\end{equation}
where we have defined the constant $\mathcal{K} = (\rho_{d0} - \mathcal{C})/\mathcal{C}$. This expression is what we expect for the evolution
of a unified fluid: a piece that redshifts with the scale factor as $a^{-3}$, as a dark matter component does, and a piece that remains constant, as vacuum energy.
This result has been found elsewhere in the literature, as in \cite{Avelino03}, which investigated the $\Lambda$CDM limit of a Chapliying gas, 
or as in the study of barotropic fluids with constant speed of sound \cite{Beca2007,Balbi07,Quevedo,Linder09}. In fact, using the language of E. Linder and R. Scherrer \cite{Linder09}, 
in our case the term proportional to $a^{-3}$ is the {\it aether} piece of a barotropic fluid. 

Now, the equation of state parameter of the dark fluid (\ref{EOSdf}) becomes
\begin{equation}
 w_d = - \frac{1}{1+ \mathcal{K} a^{-3}}.  \label{EOSdf2}
\end{equation}
We want to stress that the energy density of the dark fluid is proportional to the inverse of its equation of state parameter 

\begin{equation}
\rho_d = -  \frac{\rho_{d 0}}{(1 + \mathcal{K})}\frac{1}{w_d},    \label{rorcdm}
\end{equation}
and that its pressure, expressed in terms of the constant $\mathcal{K}$ instead of $\mathcal{C}$, is 
\begin{equation}
P_d =  -  \frac{\rho_{d 0}}{1 + \mathcal{K}}.  \label{rorcdm2}
\end{equation} 
In order to ensure the positivity of the energy density at all times, $\mathcal{K}$ must be a positive number. This implies that the pressure is
negative, a quality that allows the dark fluid to accelerate the Universe, and as we have outlined in the last section  it could take values of the order of
the critical density $(\sim 3 H_0^2/  8 \pi G)$ without affecting the behavior of the dark fluid as dark matter in astrophysical scenarios. 

The Friedmann equation (\ref{s1}) becomes

\begin{equation}
 H^2 = \frac{8 \pi G}{3} \left(\frac{\rho_{d0}}{1+\mathcal{K}} + \frac{\mathcal{K} \rho_{d0}}{1+\mathcal{K}} a^{-3} + 
                \rho_{b0} a^{-3} + \rho_{\gamma 0} a^{-4} \right). \label{s1_2}
\end{equation}
This is the same evolution equation for the scale factor as in the $\Lambda$CDM model.
This is not an accident, if one assumes $\Lambda$CDM  as the valid model and considers the total energy density
of the dark components, $\rho_T = \rho_{DM} + \rho_{\Lambda}$, the total equation of state parameter, $w_T$, defined by 
\begin{equation}
 w_T \equiv \frac{ \sum_a  w_{a} \rho_{a}}{\sum_a \rho_a}, \label{EOST}
\end{equation}
where the subindex $a$ runs over dark matter (DM) and cosmological constant ($\Lambda$), is given by 
\begin{equation}
 w_T = -\frac{1}{1+\frac{\Omega_{DM}}{\Omega_{\Lambda}} a^{-3}}, \label{EOST2}
\end{equation}
where $\Omega_{\text{DM}}  = 8 \pi G \rho_{\text{DM}0}  / 3H^2_0$ and $\Omega_{\Lambda0}  = \Lambda  / 3H^2_0$. 
(Clearly, the equation $\dot{\rho}_T = -3 \mathcal{H} (1+w_T)\rho_T$ is accomplished 
for the total energy density.) Then, comparing these results to Eqs. (\ref{EOSdf2}) and (\ref{s1_2}), we note that under the identifications 
\begin{equation}
 \mathcal{K} = \frac{\Omega_{DM}}{\Omega_{\Lambda}}, \label{deg1rel}
\end{equation}
and
\begin{equation}
 \Omega_{d} = \Omega_{DM} + \Omega_{\Lambda}, \label{deg2rel}
\end{equation}
where $\Omega_{d} = 8 \pi G \rho_{d0} / 3H^2_0$, the resulting evolution cosmology in both models are exactly the same, at least at the background level.

This property has been called {\it dark degeneracy} by M. Kunz in \cite{Kunz09}. In fact, it is more general than for the single fluid case worked here:
any  collection of fluids whose total equation of state parameter is equal to  Eq. (\ref{EOST2}) and that do not interact with 
baryons and photons will behave exactly as the composed dark matter-cosmological constant fluid, leading to a degeneracy with the $\Lambda$CDM model.

\section{Beyond the Background}

In this section we explicitly show  that the degeneracy is preserved when one goes beyond the homogeneous and
isotropic cosmology, but  only under some general assumptions that have been taken for granted in previous works.  
We divide the discussion into linear and higher orders in cosmological perturbation theory.

\subsection{Linear order}
Let us consider cosmological perturbation theory in the conformal Newtonian gauge, the metric is given by (for details see \cite{Ma_Be})
\begin{equation}
 ds^2 = a^2(\tau)\big[ -(1+2 \Psi) d\tau^2 + (1- 2 \Phi) \delta_{ij} dx^i dx^j \,\big],
\end{equation}
where $\tau$ is the conformal time related to the cosmic time by $dt = a d \tau$.
As usual, we define the matter perturbation variables through the expressions of the energy momentum tensor

\begin{eqnarray}
T^{0}{}_{0} &=& -\rho(1+ \delta), \\
T^{i}{}_{0} &=& - (\rho + P) v^i, \\
T^{i}{}_{j} &=& P \big( (1+\pi_L) \delta^{i}{}_{j} + \Pi^{i}{}_{j} \big),
\end{eqnarray}
where  $\Pi^{i}{}_{j}$ is the  anisotropic (traceless) stress tensor. The energy density $\rho$ and the pressure $P$ denotes background quantities and are 
functions of the conformal time only. The vector $v^i$ is called the peculiar velocity and is related to the four velocity $u^{\mu}$ of the fluid by the 
relation $v^i = u^i / u^0$. In Fourier space we define the velocity $\theta = i k_i v^i$ and the scalar anisotropic  stress
$\sigma = 2 k_i k_j \Pi^{ij} w / 3 (1+w)$. We use the flat space metric
$(\delta_{ij})$ to raise and lower indices of intrinsic space geometrical objects like $v^i$ and $\Pi^{i}{}_{j}$.

The hydrodynamical equations for a general fluid  are \cite{Ma_Be}

\begin{eqnarray}
 \dot{\delta} &=& - (1+ w)(\theta - 3 \dot{\Phi}) - 3 \mathcal{H} \left( \frac{\delta P}{\delta \rho} - w \right) \delta,  \label{pertd} \\
 \dot{\theta} &=& - \mathcal{H} (1 - 3 w)\theta - \frac{\dot{w}}{ 1+w} \theta + \frac{ \delta P / \delta \rho}{1+w}k^2 \delta \nonumber\\
              & &  + k^2 \Psi  -  k^2 \sigma,  \label{pertt}
\end{eqnarray}
where  $\delta P = P \pi_L$, $\delta \rho = \rho \delta$, $\mathcal{H} = \dot{a}/a$ and a dot means derivative with respect to conformal time.
For the dark fluid case, $c_s^2 = \dot{P}_d / \dot{\rho}_d = w_d - \dot{w}_d/3 \mathcal{H}(1+w_d) =0$ and these equations become

\begin{eqnarray}
 \dot{\delta}_d &=& - (1+ w_d)(\theta_d - 3 \dot{\Phi}) + 3 \mathcal{H}  w_d \delta_d \nonumber\\ 
                & &   - 3 \mathcal{H} \frac{\delta P_d}{\delta \rho_d} \delta_d,  \label{pertddf} \\
 \dot{\theta}_d &=& - \mathcal{H} \theta_d  + k^2 \Psi + \frac{\delta P_d / \delta \rho_d}{1+w_d} k^2 \delta_d -  k^2 \sigma_d,  \label{perttdf}
\end{eqnarray}
while for baryons after recombination (when the coupling to photons can be safety neglected)
\begin{eqnarray}
 \dot{\delta}_b &=& -\theta_b + 3 \dot{\Phi},  \label{pertdb} \\
 \dot{\theta}_b &=& - \mathcal{H} \theta_b  + k^2 \Psi.  \label{perttb}
\end{eqnarray}
The fluid equations are supplemented with the Einstein's equations

\begin{equation}
 k^2 \Phi = -4 \pi G a^2 \sum_i \rho_i \Delta_i, \label{Ppe}
\end{equation}
and 
\begin{equation}
 k^2( \Phi - \Psi) = 12 \pi G a^2 \sum_i (\rho_i + P_i) \sigma_i
\end{equation}
where the sum runs over all fluid contributions and 
\begin{equation}
 \Delta_i = \delta_i + 3 \mathcal{H} (1+w_i) \frac{\theta_i}{k^2} 
\end{equation}
is the rest fluid energy density \cite{Bardeen80}.

To solve these equations, we need to add information about the nature of the dark fluid. 
The barotropic condition implies that $\delta P = c^2_s \delta \rho$ and thus $\delta P_d =0$, and because it is a perfect fluid, the anisotropic stress vanishes, 
$\Pi_d^{\,\,i}{}_{j} = 0$. Accordingly, the space-space components of the perturbed energy momentum tensor are equal to zero, $\delta T^{ij}_d = 0$. Therefore, in the 
right-hand side (rhs) 
of Eq. (\ref{pertddf}) the last term vanishes, and in the rhs of Eq. (\ref{perttdf}) only the two first terms survive. If there are only baryons and
dark fluid the two gravitational potentials are equal, $\Psi = \Phi$.

\begin{figure}
\label{fig1}
\begin{center}
\includegraphics[width=3in]{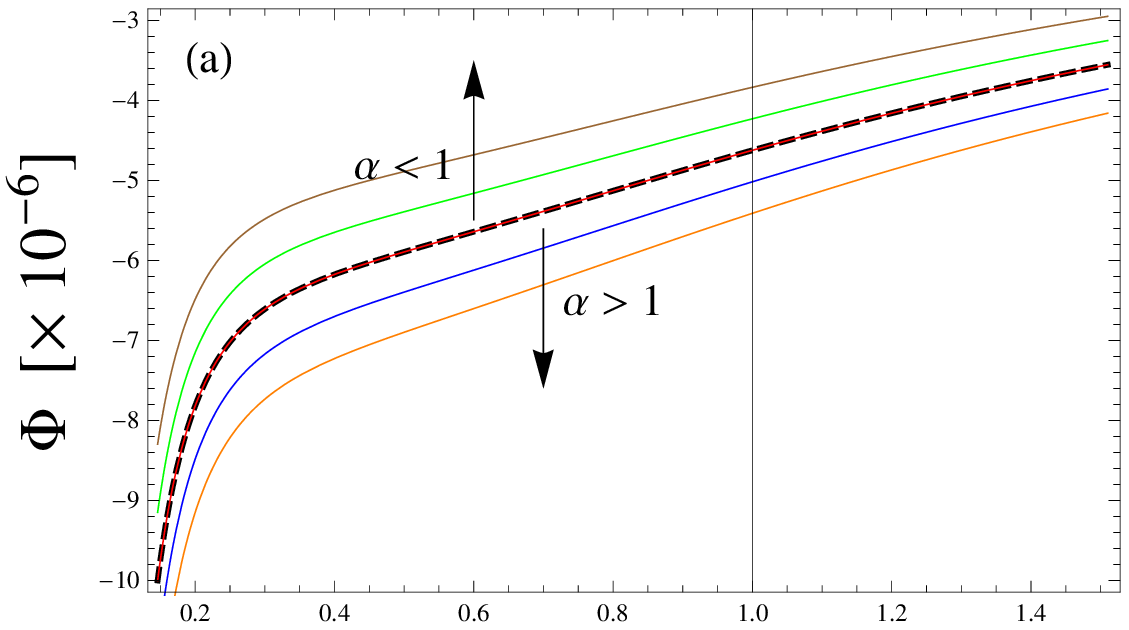} 
\includegraphics[width=3in]{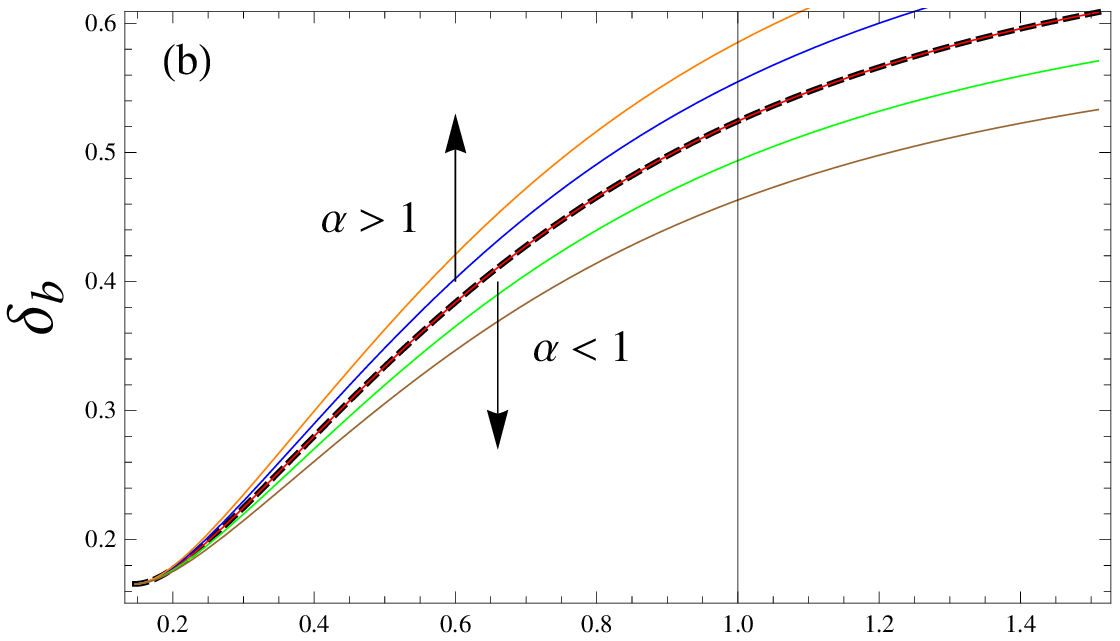}
\includegraphics[width=3in]{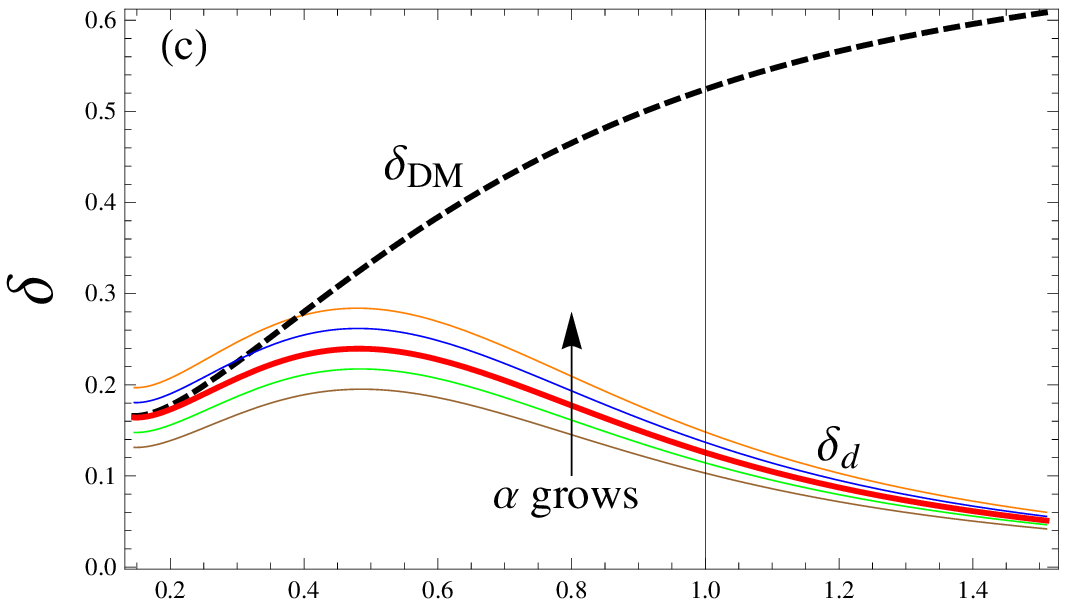}
\includegraphics[width=3in]{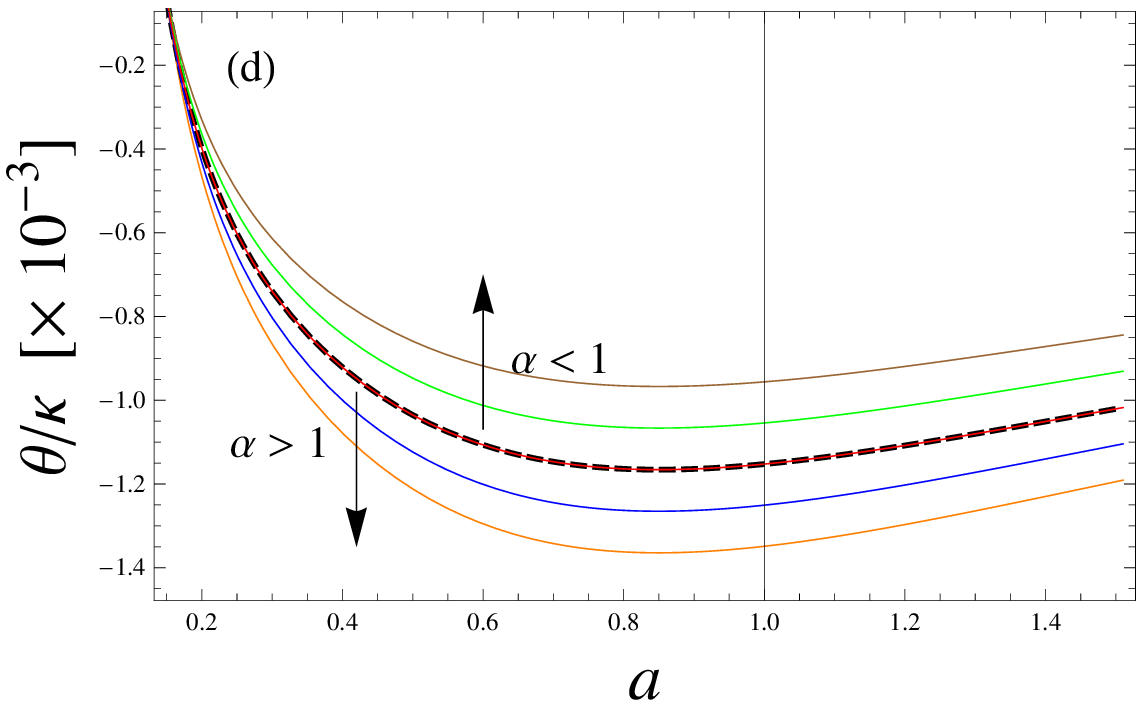}
\caption{Evolution of perturbation variables for a mode $k = 0.05\,\text{Mpc}^{-1}$. 
         Solid curves are obtained from the dark fluid model for different values of the parameter $\alpha$ in the initial conditions 
          $\delta_d(\tau_i) = \alpha \rho_{\text{DM}}(\tau_i) \delta_{\text{DM}}(\tau_i) / \rho_d(\tau_i)$. $\alpha$ takes  values from 0.8 to 1.2.
          Dashed curves are for the $\Lambda$CDM model variables. The panels show: 
          (a) Gravitational potential $\Phi$. (b) Baryonic density contrast $\delta_b$. (c) Dark fluid (solid lines) 
          and dark matter (dashed line) density contrasts, $\delta_{d}$ and  $\delta_{\text{DM}}$. (d) Dark fluid (solid lines)  
          and dark matter (dashed line) velocities, $\theta_d$ and $\theta_{\text{DM}}$.  The solutions for the case $\alpha =1$ are depicted with the thick  (red) lines, 
          which for panels (a), (b), and (d) coincide with the dashed lines.}
\end{center}
\end{figure}

Figure 1 shows the evolution of  a mode $k = 0.05\, \text{Mpc}^{-1}$ of the perturbations variables.
The dashed lines shows results for the $\Lambda$CDM model, for which we have to use the dark matter perturbations equations, 

\begin{eqnarray}
 \dot{\delta}_{\text{DM}} &=& -\theta_{\text{DM}} + 3 \dot{\Phi},  \label{pertddm} \\
 \dot{\theta}_{\text{DM}} &=& - \mathcal{H} \theta_{\text{DM}}  + k^2 \Psi,  \label{perttdm}
\end{eqnarray}
instead of Eqs. (\ref{pertddf}) and (\ref{perttdf}).  

In evolving the perturbations for both models, $\Lambda$CDM and dark fluid, we have imposed on the initial conditions the relations  $\delta_d(\tau_i) = \alpha  \rho_{\text{DM}}(\tau_i) 
\delta_{\text{DM}}(\tau_i) / \rho_d(\tau_i)$ for the density contrasts, and $\theta_d(\tau_i) = \alpha  \rho_{\text{DM}}(\tau_i) \theta_{\text{DM}}(\tau_i) / (1+w_d)\rho_d(\tau_i)$ 
for the velocities, and we let $\alpha$ to take different values. These initial conditions are given at 
an initial time  well after recombination, so we can use Eqs. (\ref{pertdb}) and  (\ref{perttb}) and the relation $\delta_{\text{DM}} \simeq \delta_b$ holds. 

We note that the evolution of the density contrast of baryons for the specific mode we have chosen is
indistinguishable for both models if we take $\alpha=1$. Then, although the cosmological observable is the baryonic matter power spectrum which
includes a wide range of wavelengths, Fig. 1 suggests that indeed the results are the same for both models for any wavelength, as we show below.

 In the cosmological  context, imposing these two initial 
conditions is equivalent to demand  
that at first order in perturbation theory, the time-time and time-space components of the perturbed energy momentum tensor of the dark fluid and $\Lambda$CDM models
are equal  at the given initial time $\tau_i$.  It is straightforward to show that both conditions, 
in the case $\alpha=1$, will be preserved at all times. This means that in obtaining the $\alpha=1$ solutions in Fig. 1 we have imposed that


\begin{equation}
 \rho_d \delta_d = \rho_{\text{DM}} \delta_{\text{DM}}, \label{ass1}
\end{equation}
and
\begin{equation}
 \rho_d(1+w_d) \theta_d = \rho_{\text{DM}} \theta_{\text{DM}}.  \label{ass2}
\end{equation}
From now on, we will demand Eqs. (\ref{ass1}) and (\ref{ass2}) to hold for all considered fluids, staying in
degeneracy with the $\Lambda$CDM model   at first order in cosmological perturbation theory. Later on, in Sec. VI, we 
will try to break this degeneracy, but through couplings of the dark fluid to baryonic matter.

From the results of the previous section it is straightforward 
to see that $\rho_{\text{DM}}/\rho_d = 1+w_d$, then Eq. (\ref{ass2}) implies that $\theta_{\text{DM}} = \theta_d$, 
which explains the behavior of the curves  in Fig. 1d. Moreover, from Eq. (\ref{Ppe}) it follows that the gravitational potentials $\Phi$ are the same
for both models, so the behavior in Fig. 1a. 

Finally,  inserting  Eqs. (\ref{ass1}) and (\ref{ass2}) into Eqs. (\ref{pertddm}) and  (\ref{perttdm}), respectively, and using Eq. (\ref{rorcdm}),  we obtain that the 
hydrodynamical perturbation equations for the dark fluid are 

\begin{eqnarray}
 \dot{\delta}_d &=& - (1+ w_d)(\theta_d - 3 \dot{\Phi}) + 3 \mathcal{H}  w_d \delta_d,
                              \label{pertddf2} \\
 \dot{\theta}_d &=& - \mathcal{H} \theta_d  + k^2 \Psi.  \label{perttdf2}
\end{eqnarray}
These equations are equal to (\ref{pertddf}) and (\ref{perttdf}) for a barotropic perfect fluid, confirming
the dark degeneracy at first order level. 

It is interesting to note the behavior of the density contrast of the dark fluid (Fig. 1c), which
initially grows as a dark matter component to later decay asymptotically to zero. In fact, this behavior is natural for all wavelengths modes as can 
be seen from Eqs. (\ref{pertddf}) and (\ref{perttdf}): At late times, when $w_d$ tends to -1, the $\dot{\delta}_d$ source becomes independent of 
$\theta_d$ and is a negative quantity which tends to zero. Meanwhile at early times, for   
$w_d \rightarrow 0$, $\mathcal{H}w_d \rightarrow 0$, and the equations become equal to the cold dark matter evolution equations, (\ref{pertddm}) and (\ref{perttdm}).

\subsection{Beyond linear order}

To go beyond the linear order, let us make  perturbation expansions to the dark fluid  and $\Lambda$CDM
energy  momentum tensors about the (zero order) background cosmological fluids as 
\begin{equation}
 T_{\mu\nu} =  T_{\mu\nu}^{(0)} + T_{\mu\nu}^{(1)} +T_{\mu\nu}^{(2)} + \cdots. \label{EMTexp}
\end{equation}
If the total energy momentum tensors of both models are equal $\,\,\big(T_{\mu\nu}^{\,d} = T_{\mu\nu}^{\Lambda\text{CDM}}\big)$, clearly each of the terms 
in the expansion will be equal as well $\big(T_{\mu\nu}^{\,d\,(i)} = T_{\mu\nu}^{\Lambda\text{CDM}\,(i)} \big)$. This  argument is  outlined in \cite{Kunz09}
to argue that the degeneracy is preserved at all orders. Certainly, this is correct. 

Nonetheless, we want to stress a different approach: We are affected gravitationally by the total energy momentum tensor, but usually
when comparing observations to models we expand it as in Eq. (\ref{EMTexp}), and after this we assign values to each of the pieces. 
The fact that both energy momentum tensors are equal, say at zero order, does not imply that they will be equal at first order. 
In this situation, equations such as (\ref{ass1}) and (\ref{ass2}) are conditions of the theory and not consequences of it, and if not imposed, 
the degeneracy is broken, as seen in Fig.1 for $\alpha \neq 1$. 

This approach, which we will adopt, is very close to that followed in some papers which investigate modified gravity theories
that are indistinguishable in the background from the $\Lambda$CDM model but could differ in linear order,  leaving imprints that are parametrized to compare to 
current and future observations (see \cite{Daniel10} and references therein). 
In fact, if some of these imprints were detected, they could be a consequence of interactions within the dark
sector (or even with baryons, as those we study in Sec. VI), instead of a deviation of General Relativity.

Finally, there is a subtle point that is pertinent to assert at this moment. The fact that the Universe is so smooth at very large scales is what 
allows us to expand the energy momentum tensor of the cosmic fluids as in Eq. (\ref{EMTexp}). The zero-order term is in fact a spatial average in hypersurfaces 
of constant cosmic time, $T_{\mu\nu}^{(0)} = \langle T_{\mu\nu} \rangle$, and thereafter we assume that general relativity holds in the form 
$G_{\mu\nu} (g^{(0)}) = \langle T_{\mu\nu} \rangle$. 
This is not true in general due to the nonlinear character of gravity; it is well known that backreaction effects contributes to this last equation. 
For recent work in the averaging procedure in cosmology and its caveats see \cite{Green2011}.

For the specific case that concerns us there is a related issue. Let us consider a fluid  with equation of state $P = P(\rho)$ that leads to a unified  description of the dark sector. 
It has been noted  that it is not necessarily true that $\langle P \rangle = P(\langle \rho \rangle)$ \cite{Avelino2004}; in fact higher-order corrections enter into this equation. 
Therefore, when some scale grows 
and leaves the linear regime, the naive averaging procedure on the equation of state is no longer valid. This nonlinear effect has been investigated in 
\cite{Beca2007,Avelino2004,Avelino2008}, and it was shown that instabilities are expected to occur in unified dark models, even on large cosmological scales. 
Two exceptions are the cases in which the involved equations of state are $ P = \text{constant}$ and $ P= w \rho $, 
with $w$ a constant. The former is the case of the dark fluid, and then, it does not suffer for this {\it averaging problem}. Nonetheless, it is worth  
emphasizing that in models that depart --although indistinguishable by current observations when calculations are performed using the average procedure-- from the dark fluid 
or from the $\Lambda$CDM models these effects must be considered.

\section{Multiple Interacting Dark Fields}

From the last section it is clear that the dark fluid, although not necessarily, could be the sum of a dark matter and a cosmological constant components. 
Nonetheless, if the interaction to the particles of the standard model is only gravitational (which is the ultimate definition of {\it dark}), 
the nature of the dark fluid is fundamentally impossible to elucidate, because of the universality of this force. 

In this section we explore the possibility that
the dark fluid is composed of a collection of dark components with possible interactions between them. The equations that we obtain will be generalized 
 in the next section to include interactions to baryons. Let us consider the Einstein field equations,

\begin{equation}
 G^{\mu\nu} = 8 \pi G \left( \sum_a T^{\mu\nu}_{a} + T^{\mu\nu}_{\text{SM}} \right),
\end{equation}
where the subindex $a$ labels the different dark components. 
If we forbid dark components to interact with standard model particles, the Bianchi identities imply $\nabla_{\mu} T^{\mu\nu}_{\text{SM}} = 0$ and

\begin{equation}
 \nabla_{\mu} T^{\mu\nu}_{a} = Q^{\nu}_{a}, \label{Biid}
\end{equation}
where the energy momentum transfer vectors, $Q^{\nu}_{a}$, obey the constraint 
\begin{equation}
 \sum_a Q^{\nu}_{a} = 0. \label{condQ}
\end{equation}
Considering for the moment the homogeneous and isotropic cosmology, the continuity equation for each fluid is

\begin{equation}
 \dot{\rho}_a +3 \mathcal{H}(1 + w_a)\rho_a = q_a,
\end{equation}
where $q_a$ denote the energy transfer between different dark components. This implies that at zero order, (ignoring the subindex $a$'s for the moment) 

\begin{equation}
Q^{\nu\,(0)} = \frac{1}{a^2}(q,\vec{0}),
\end{equation}
if we define  $Q \equiv \sqrt{-g_{\mu\nu}Q^{\mu}Q^{\nu}}$, then $Q^{(0)}=q/a$ and $ Q^{\nu\,(0)} = a^{-1}(Q^{(0)},\vec{0}).$
Now, following \cite{Kodama,MalikWands}, we split the interaction vector as  
\begin{equation}
 Q_{\nu} = Q u_{\nu} + f_{\nu}, \label{descomQ}
\end{equation}
where the momentum transfer, $f^{\nu}$, is first order and is orthogonal to the four velocity $f^{\nu}u_{\nu}=0$. 
Thus, under spatial rotations $Q$ transforms as a scalar and $f^{i}$ as a vector. Accordingly, we decompose $f^i$ as
\begin{equation}
 f_i = f_{,i} + \epsilon_i 
\end{equation}
with $f$ a scalar and $\epsilon_i$ a transverse vector, $\epsilon^i_{\,,i}=0$, and 
we define the perturbation $\delta Q$ through
\begin{equation}
Q = Q^{(0)}+ \delta Q = \frac{1}{a}( q+\delta q),
\end{equation}
where in the last equality we defined  $\delta q \equiv a \delta Q$.
From (\ref{descomQ}) it follows that up to first order

\begin{eqnarray} 
 Q^{0} &=&  \frac{1}{a^2} \big( q(1-\Psi) + \delta q \big)  \label{Q0} \\
  Q^{i} &=& \frac{1}{a^2} q v^i + \frac{1}{a^2} f^{,i} + \frac{1}{a^2} \epsilon^i. \label{Qi}  
\end{eqnarray}
Finally, from Eq. (\ref{condQ}) one obtains (after restoring $a$'s) the constriction equations
\begin{equation}
 \sum_{a} \delta q_{a} = 0, \qquad\qquad \sum_{a} q_{a} = 0, \label{constf2}
\end{equation}
and $a^2 \sum_{a} Q^i_{a} = \sum_{a}  (q_{a} v^i_{a} + f^{,i}_{a} +  \epsilon^i_{a})$, which by
 taking the divergence and going to Fourier space become

\begin{equation}
 \sum_{a} \left(q_{a} \theta_{a} + k^2 f_{a} \right) =0.   \label{constf}
\end{equation}
Now, to first order in perturbation theory, the divergence of the energy momentum tensor --for each fluid, omitting the subindex $a$-- is

\begin{eqnarray}
 \nabla_{\mu}T^{\mu 0} &=& \frac{1}{a^2}(1+\delta - 2 \Psi)(\dot{\rho} + 3\mathcal{H}(1+w)\rho) \nonumber\\
   & + &  \frac{\rho}{a^2} \Bigg[ \dot{\delta} + (1+w)(\partial_i v^i - 3 \dot{\Phi}) \nonumber\\ 
   &  &  \qquad+ \,3\mathcal{H}\left( \frac{\delta P}{\delta \rho} - w \right) \delta   \Bigg] \label{emtc0}, 
\end{eqnarray}

\begin{eqnarray}
 \nabla_{\mu}T^{\mu i} &=& \frac{P}{a^2} \partial_j \Pi^{ij}  + \frac{1}{a^2} (\dot{\rho} + 3\mathcal{H}(1+w)\rho)v^i \nonumber\\
 &+ & \frac{\rho + P}{a^2} \Bigg[ \dot{v}^i + 
\mathcal{H} \left(1+\frac{\dot{\rho} w}{\mathcal{H}(1+w)\rho} \right) v^i \nonumber\\ 
& & \qquad + \frac{\dot{w}}{1+w} v^i + \frac{w \pi_L^{,i}}{1+w} + \Psi^{,i} \Bigg] \label{emtci}.
\end{eqnarray}
Note that if the interaction vanishes, both equations reduce to the usual ones. 
We take the divergence of Eq. (\ref{emtci}) to isolate the scalar mode perturbations, then Eq. (\ref{Biid}) implies

 \begin{eqnarray}
 \dot{\delta}_a &+& (1+w_a)(\theta_a - 3 \dot{\Phi}) + \nonumber\\
 3 \mathcal{H} \left( \frac{\delta P_a}{\delta \rho_a} - w_a \right)\delta_a
&+&\frac{q_a}{\rho_a} (\delta_a - \Psi ) - \frac{\delta q_a}{\rho_a} \,=\, 0
 \end{eqnarray}
and

\begin{eqnarray}
\dot{\theta}_a +  \mathcal{H} \Bigg(1\!\!&-&\!\! 3w_a +\frac{q_a\, w_a}{\mathcal{H}(1+w_a)\rho_a} \Bigg) \theta_a \nonumber \\ + 
\frac{\dot{w}_a}{1+w_a} \theta_a  
\!\!&\!\!-\!\!&\!\!  \frac{\,\delta P_a / \delta \rho_a \,}{1+w_a}k^2 \delta_a - k^2 \Psi \nonumber\\
+ k^2 \sigma_a \!\!&\!\!+\!\!&\!\! \frac{k^2 f_a}{\rho_a(1+w_a)}  = 0,
\label{eqti2}
\end{eqnarray}
where we have restored the index $a$'s and back to Fourier space. Also we have used the decomposition for $\Pi^{ij}$ in tensor, vector and scalar pieces.
Then when hit with $k_i k_j$ only the scalar anisotropic stress survives
and the first term of the rhs of Eq. (\ref{eqti2}) becomes $2 k^2 \Pi^{(s)} P /3= (\rho + P) k^2  \sigma $.

Defining the total energy density of these dark fluids as $ \rho_T = \sum \rho_a$,
it follows that the total density contrast, $\delta_T$, is the weighted sum of the individual components
\begin{equation}
 \delta_T = \frac{1}{\rho_T} \sum_a \rho_a \delta_a,
\end{equation}
and the total velocity is given by
\begin{equation}
 \theta_T = \frac{1}{\rho_T (1+w_T)} \sum_a \rho_a(1+w_a) \theta_a,
\end{equation}
where $w_T$ is given by Eq. (\ref{EOST2}), and the index $a$ runs over all dark components.
Using these two last equations it is straightforward to show that the hydrodynamical equations for the total fluid  are

\begin{equation}
 \dot{\delta}_T + (1+w_T)(\theta_T - 3 \dot{\Phi}) + 3 \mathcal{H} \left(\frac{\delta P_T}{\delta \rho_T}- w_T \right)\delta_T =0  \label{eqdT2}
\end{equation}
and
\begin{eqnarray}
 \dot{\theta}_T &+& \mathcal{H} (1 - 3 w_T) \theta_T + \frac{\dot{w}_T}{1+w_T} \theta_T \nonumber \\
   &-& \frac{\delta P_T / \delta \rho_T}{1+w_T} k^2 \delta_T - k^2 \Psi + k^2 \sigma_T = 0,
\end{eqnarray}
where $\delta P_T = \sum \delta P_a$. These are exactly the equations of a single noninteracting fluid, Eqs. (\ref{pertd}) and (\ref{pertt}). The adiabatic 
speed of sound of the total fluid is $c^2_{sT} = \dot{P}_T / \dot{\rho}_T = \sum \rho_a c^2_{sa}/ \rho_T$. Thus, by forcing the condition $c^2_{sT} =0$ we obtain
that $w_T =w_d$, and if each dark component has positive energy ($\rho_a >0$), it also implies that  $c^2_{sa} = 0$ and  
 $\delta P_a =0$. Therefore, we obtain that  the hydrodynamical equations for the total fluid perturbations 
 become the same as the dark fluid perturbations (Eqs. (\ref{pertddf2}) and (\ref{perttdf2})), under the substitution $T \rightarrow d$. Thus, the composed fluid
we have constructed is just the dark fluid.

If one considers the unphysical case in which not all dark components have positive energy density,  the speed of sound of some components can be different from zero, and the total 
pressure perturbation is equal to 
\begin{equation}
 \delta P_T = c^2_{sT} \delta \rho_T -\frac{1}{6 \mathcal{H} \dot{\rho}_T} \sum_{a,b} \dot{\rho}_a \dot{\rho}_b (c^2_{sa}-c^2_{sb}) \mathcal{S}_{ab}, \label{dPT}
\end{equation}
where 
\begin{equation}
\mathcal{S}_{ab} = \frac{\delta_a}{1+w_a} - \frac{\delta_b}{1+w_b}, 
\end{equation}
is the relative entropy perturbation between fluids $a$ and $b$ \cite{Malik_Wands_2005}. Therefore, to obtain the dark fluid in the case in 
which some components have negative energy density we have to impose the pressure perturbations given by Eq. (\ref{dPT}) to be equal to zero. 
This condition also implies that the total fluid is barotropic.

\section{Interactions to the standard model of particles.}
In this section we study  dark fluid couplings to baryonic matter. We consider models in which the background cosmology is the same as in the $\Lambda$CDM model, 
accordingly we do not allow energy transfer ($q=0$) between the cosmic components. Nevertheless, we allow a momentum transfer different from zero. Thus, the interactions
affect the fluids only at first order in perturbation theory.

In this work we will not consider interactions to electromagnetism. This is not only for simplicity,
many theoretical models present conformal couplings, as the chameleon theories \cite{khoury1_1,khoury1_2} 
(and in general, scalar tensor gravity theories \cite{Brans_Dicke}), or even direct couplings
to the trace of the energy momentum tensor \cite{Aviles,NarPad85,SinghPad88,SamiPad03,Nicolis}. 
Also, this is expected in scenarios like the strong interacting dark matter  \cite{Spergel2000,Wandelt2001}, where the couplings are through the strong force, 
for which photons are chargeless.

\begin{figure}
\begin{center}
\includegraphics[width=3.4in]{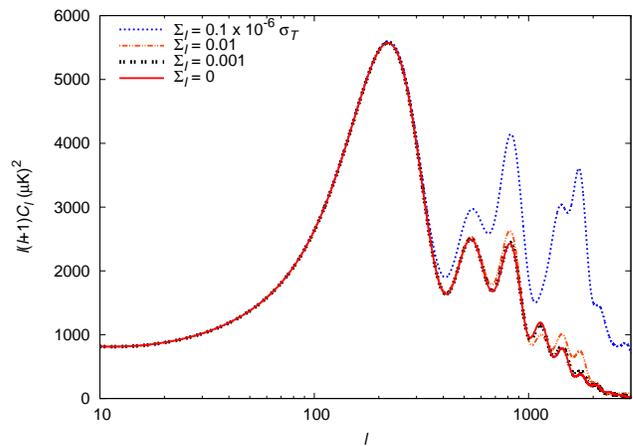} 
\caption{The CMB power spectrum considering different values of the interaction parameter $\Sigma_{I}$ in units of $10^{-6}  \sigma_T $. The other parameters
         are fixed.}
\end{center}
\label{fig2}
\end{figure}

In the previous section we found the hydrodynamical equations for a system of coupled dark components. It is straightforward to generalize those results
to a coupling between the dark fluid  and baryonic matter. 
The conservation equations for the perturbations are, for the dark fluid,
\begin{eqnarray}
 \dot{\delta}_d &=& - (1+ w_d)(\theta_d - 3 \dot{\Phi}) + 3 \mathcal{H}  w_d \delta_d + \frac{\delta q_d}{\rho_d}
                              \label{pertddfc} \\
 \dot{\theta}_d &=& - \mathcal{H} \theta_d  + k^2 \Psi - \frac{k^2 f_d}{\rho_d (1+w_d)},  \label{perttdfc}
\end{eqnarray}
and for baryons
\begin{eqnarray}
 \dot{\delta}_b &=& -\theta_b + 3 \dot{\Phi} + \frac{\delta q_b}{\rho_b},  \label{pertdbc} \\
 \dot{\theta}_b &=& - \mathcal{H} \theta_b  + k^2 \Psi + c^2_{sb} k^2 \delta_b - \frac{k^2 f_b}{\rho_b (1+w_b)}.  \label{perttbc}
\end{eqnarray}
These equations are supplemented by  the constrictions (\ref{constf2}) and (\ref{constf}). The interactions to electromagnetism are considered, but 
not shown in Eq. (\ref{perttbc}). Note that we have not considered a term $c^2_{sd} k^2 \delta_d$ in Eq. (\ref{perttdfc}), in agreement with the definition 
of dark fluid.

\begin{figure}
\begin{center}
\includegraphics[width=3.4in]{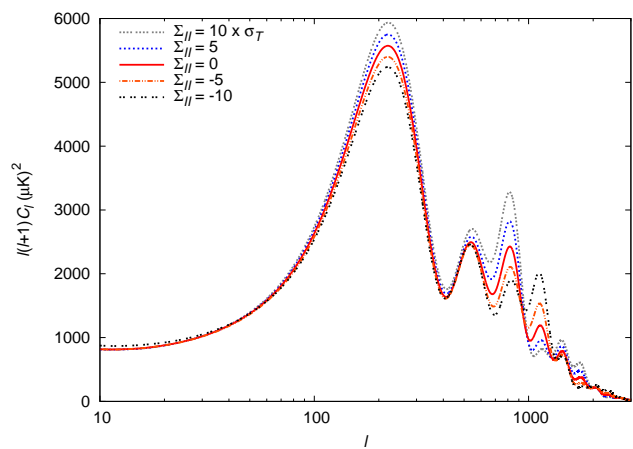} 
\caption{The CMB power spectrum considering different values of the interaction parameter $\Sigma_{II}$ in units of $\sigma_T$. The other parameters
         are fixed.}
\end{center}
\label{fig3}
\end{figure}

\begin{figure*}
\begin{center}
\includegraphics[width=2.2in]{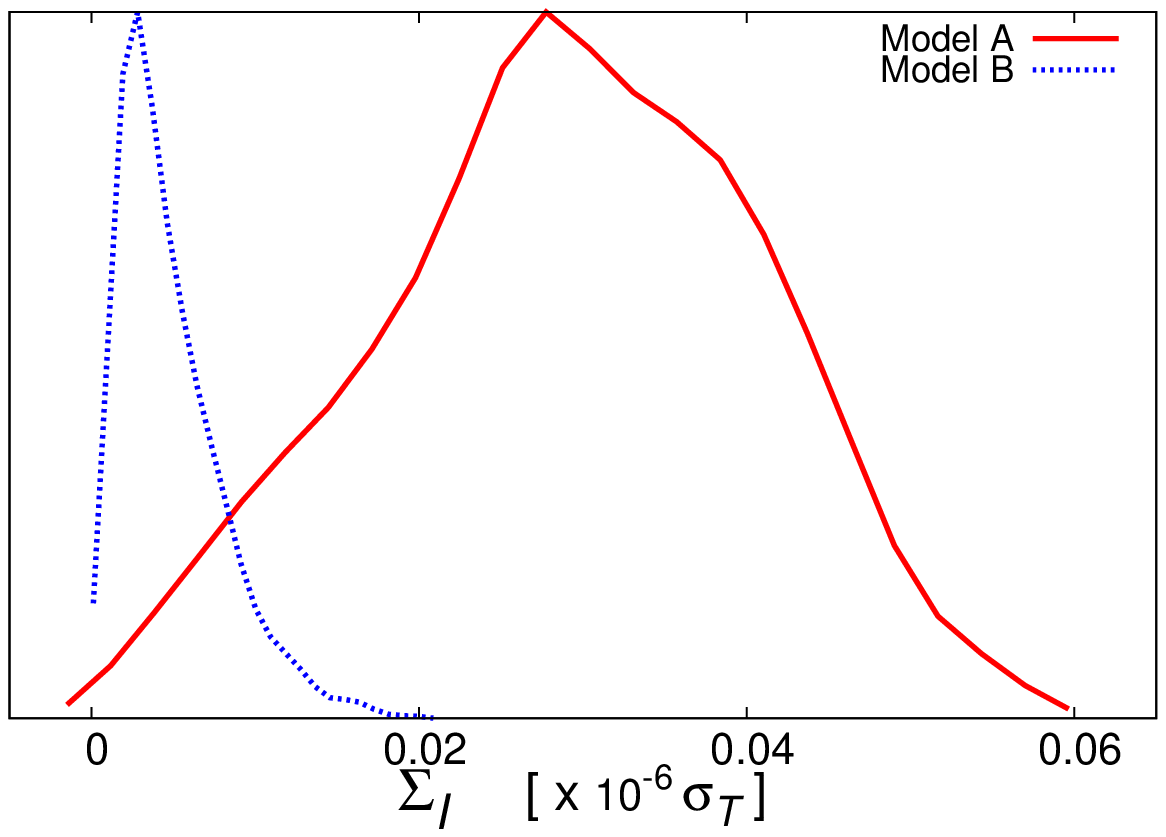}
\includegraphics[width=2.2in]{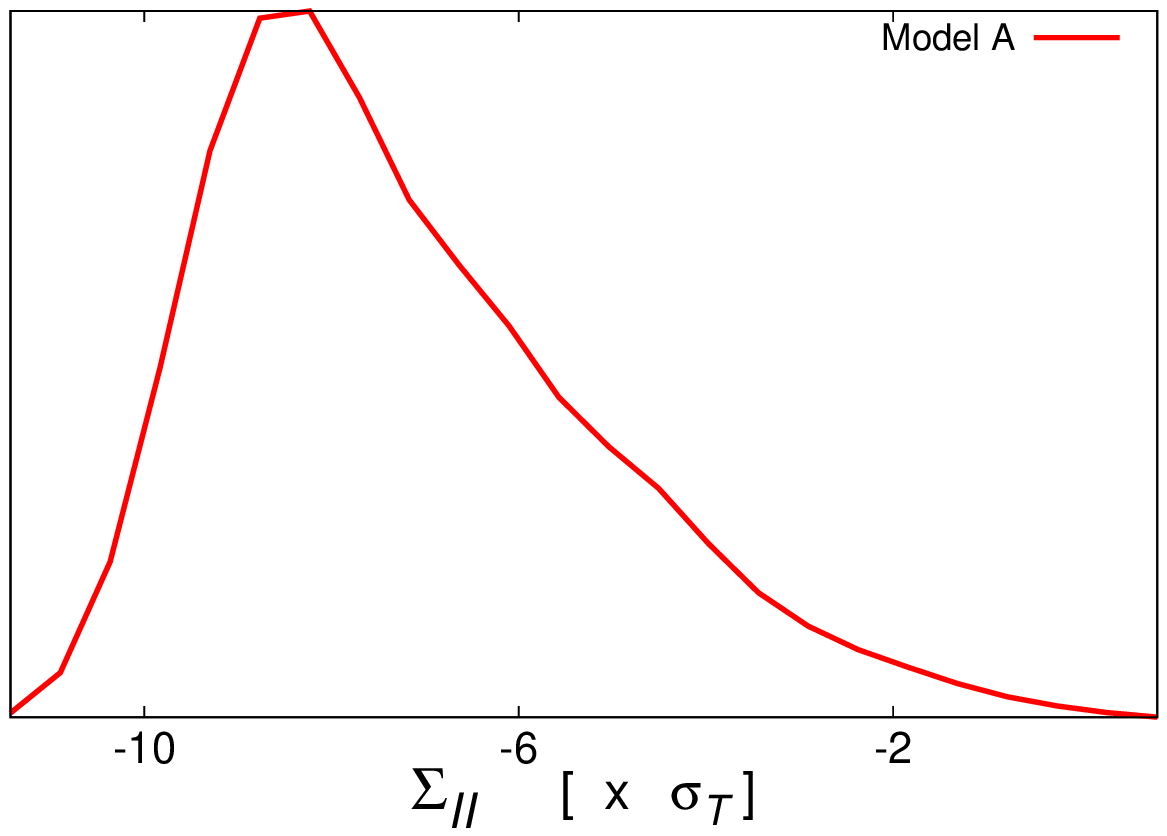}
\includegraphics[width=2.2in]{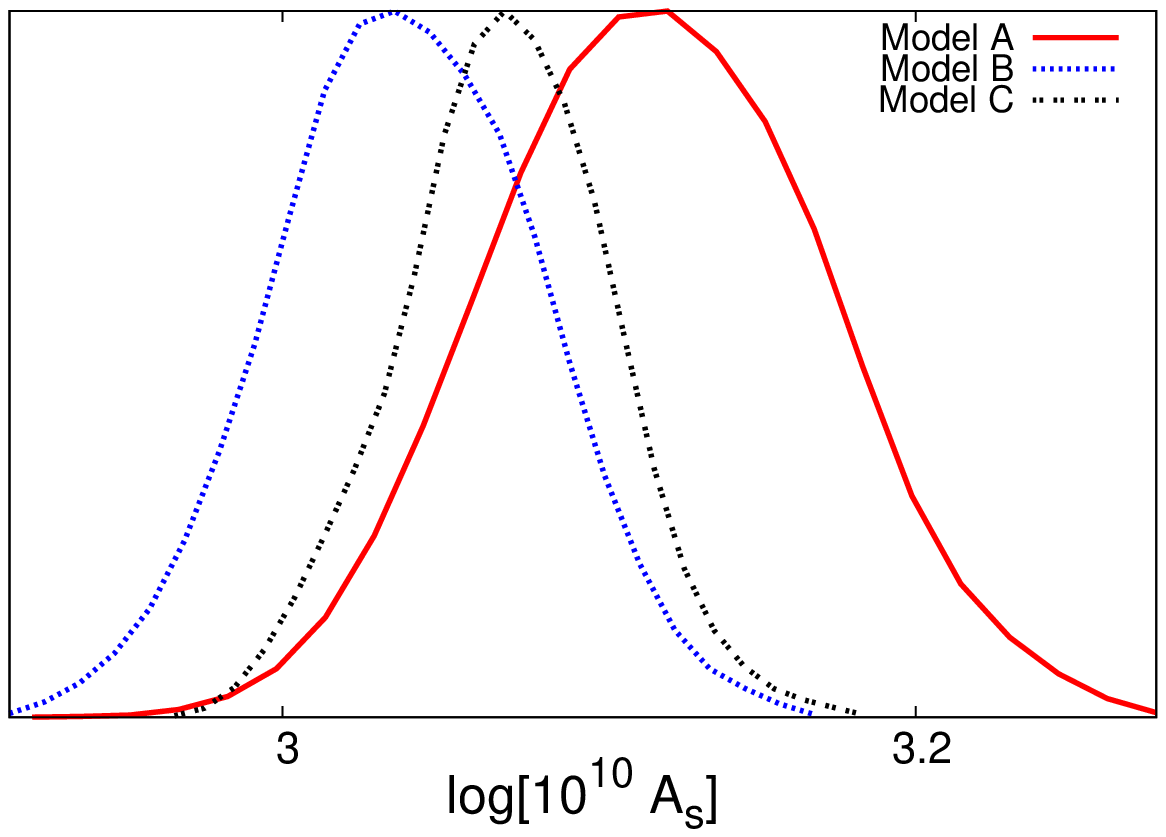}
\includegraphics[width=2.2in]{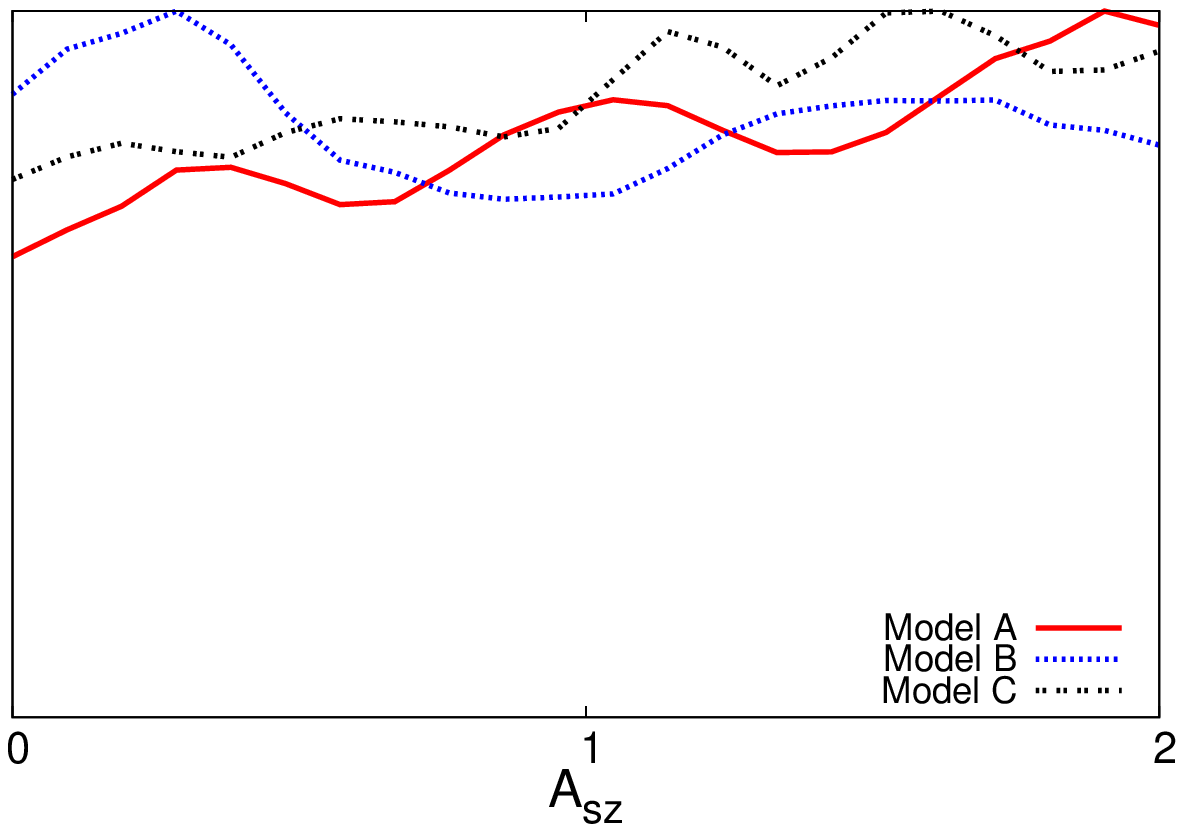}
\includegraphics[width=2.2in]{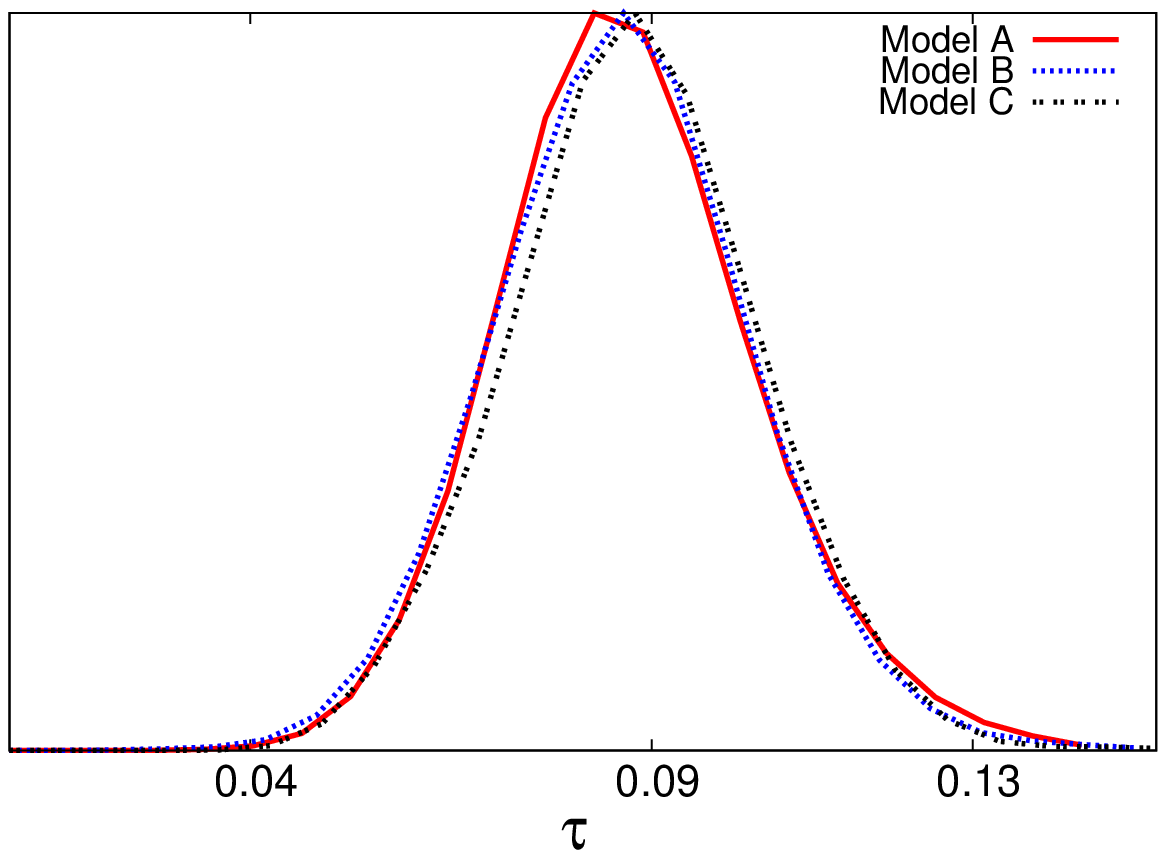}
\includegraphics[width=2.2in]{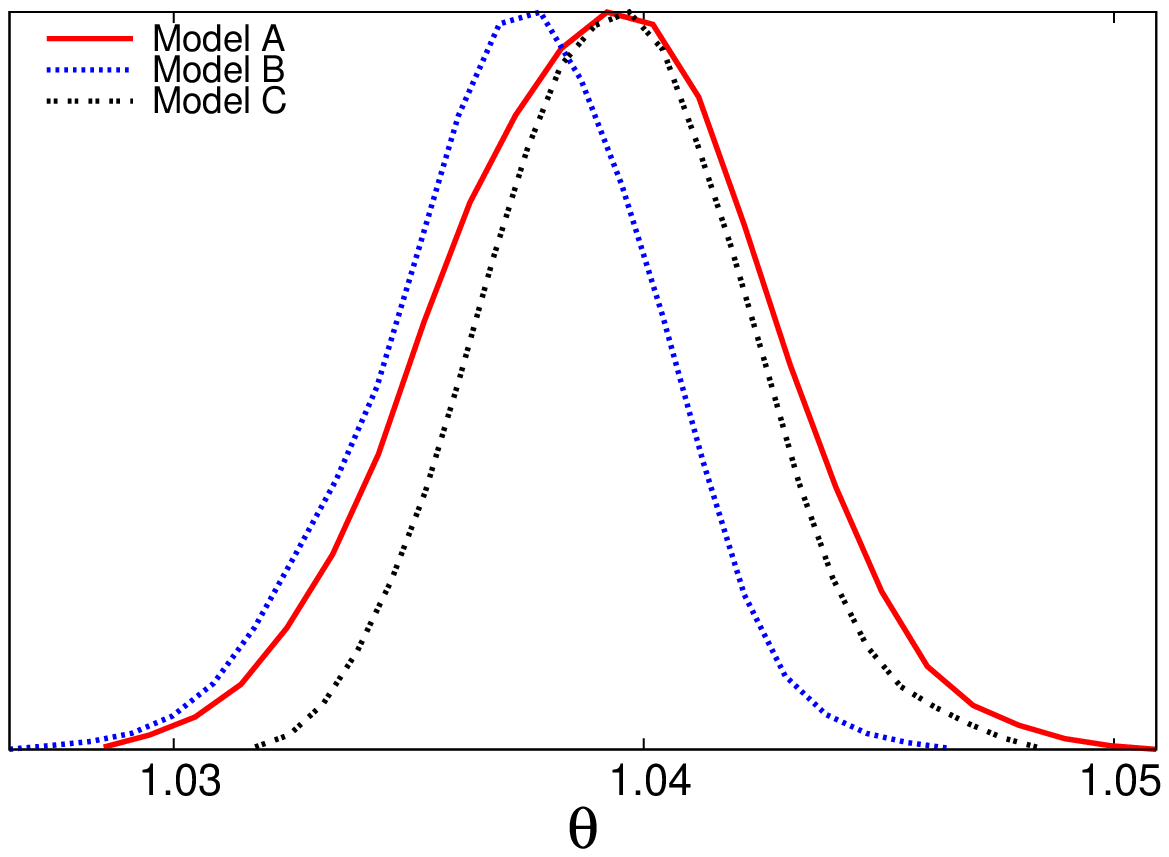}
\includegraphics[width=2.2in]{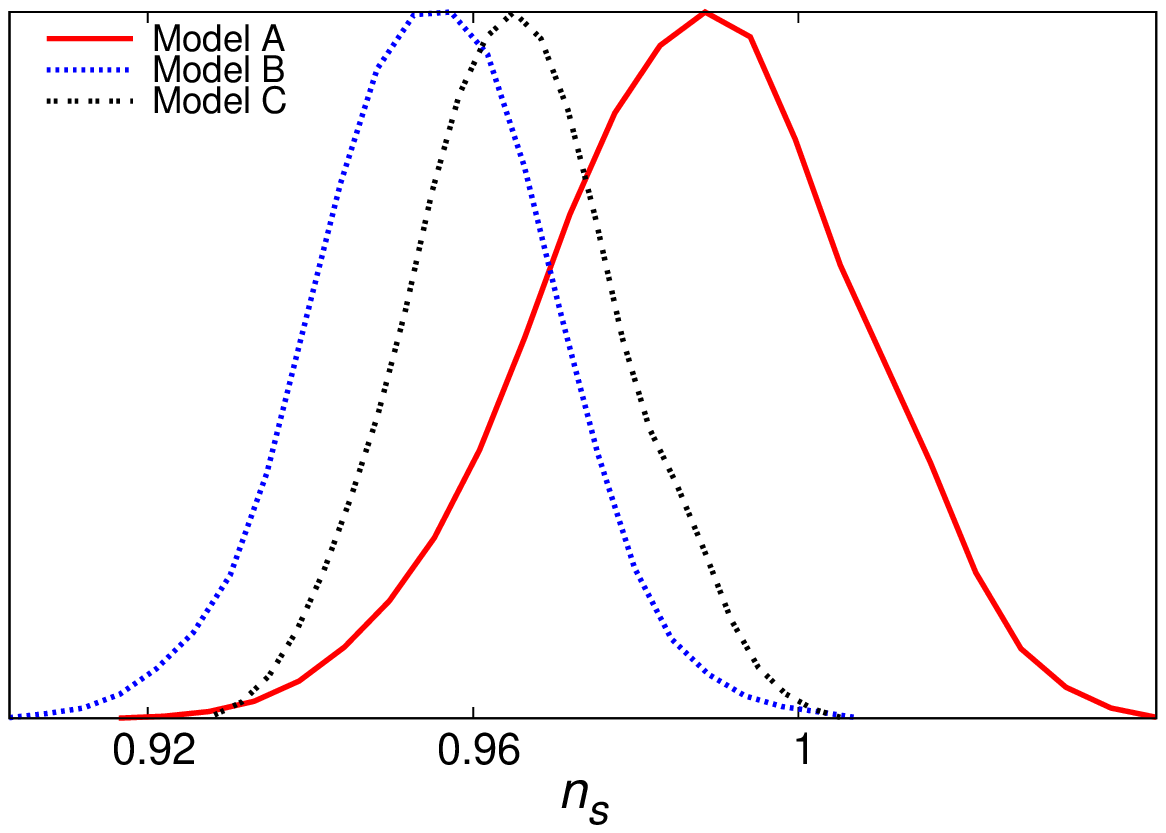}
\includegraphics[width=2.2in]{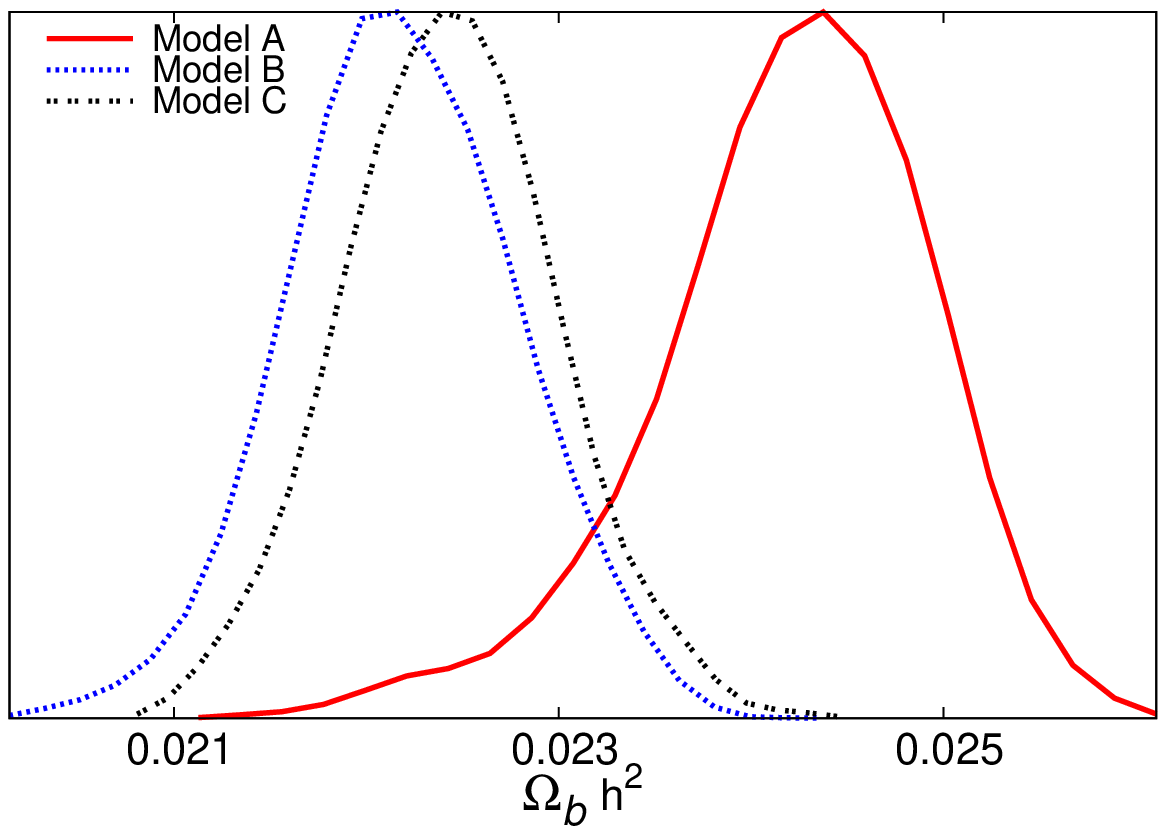}
\includegraphics[width=2.2in]{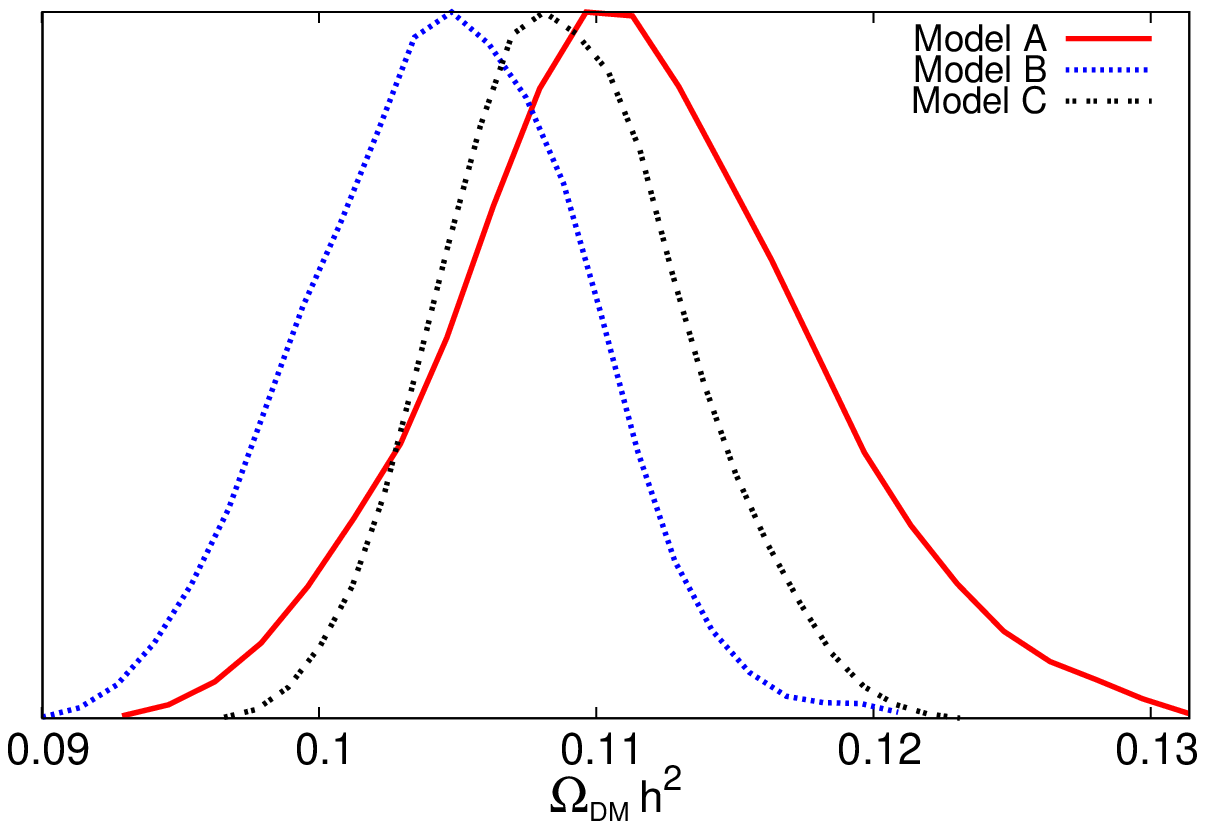}
\caption{(color online) Marginalized probability for  the complete set of parameters. Solid lines (red) are for Model A, 
dotted lines (blue) for Model B, and double-dotted (black) lines for Model C. The data used are the WMAP seven-year results, Union 2 data set supernovae 
compilation and a prior of HST on the Hubble constant.}
\end{center}
\label{fig8}
\end{figure*}

In the absence of a fundamental theory we parametrize the coupling. But before do it, let us make a comparison to electromagnetism. 
Under the formalism we developed in the last section, the hydrodynamical perturbation equations for Thomson
scattering (neglecting photon moments beyond the quadrupole) can be obtained if we choose $\delta q_{\gamma} = 0$
and $ k^2 f_{\gamma} = - \rho_{\gamma} (1+w_{\gamma}) a   x_e n_e \sigma_T c (\theta_{b} - \theta_{\gamma})$, where $\sigma_T = 6.65 \times 10^{-25} \text{cm}^2$ is the 
Thomson scattering cross section, $n_e$ is the number density of electrons,  $x_e$ is the ionization fraction, and $c = 1$ is the velocity of
light

Using the analogy to electromagnetism we choose the interactions terms to be

\begin{equation}
\delta q_d= 0, 
\end{equation}
and
\begin{equation}
 f_{d} = \rho_d(1+w_{d}) a n_d \Sigma_{I}   (\theta_{b} - \theta_d) / k^2,  \label{interaction}
\end{equation}
where the parameter $\Sigma_{I}$ has  units of area times velocity, or thermalized cross section $\langle \sigma v \rangle $, which we identify with  some, unknown, 
fundamental interaction. $n_d$ is the number density of  dark {\it particles} that we set equal to 
\begin{equation}
 n_d= \frac{\rho_{d0}}{m_p a^{3}},
\end{equation}
where we use the proton mass, $m_p = 0.938 \,\text{GeV}$, as an arbitrary mass scale and $\rho_{d0}$ is the energy density of the dark fluid evaluated today. Here, we have not an 
analogous to the ionization fraction, in empathy to universal interactions.

The constrictions $f_b=-f_d$ and $\delta q_d = -\delta q_b= 0$, enter into the baryons equations.   
Because of the relation $\rho_{DM}=(1+w_d)\rho_d$ the  interaction term  goes as $ (\theta_{d} - \theta_b) a^{-2}$. 
Then, the greatest deviation from the standard model of cosmology will come from the early Universe. Accordingly we expect the interaction to be tightly bounded 
from CMB anisotropies tests. We have modified the public available code CAMB \cite{Lewis2000} to account for the interaction.

\begin{table*}
\caption{{\small Summary of constraints. The upper panel contains the parameter spaces ex\-plo\-red with MCMC for each one of the three models. 
The bottom panel contains derived parameters. The data used are the WMAP seven-year data, Union 2 compilation and HST.}}          
\begin{tabular}{l|c|c|c} 
\hline\hline 
{\small Parameter}   & {\small Model A$\,{}^{a}$ }& {\small Model B$\,{}^{a}$} & {\small Model C$\,{}^{a}$} \\ [1.5ex]
\hline 
 {\small $10^2 \Omega_b h^2$}         & {\small $2.420$}{\tiny${}_{-0.064}^{+0.066}$}         & {\small $2.219$}{\tiny ${}_{-0.056}^{+0.056}$}        
                                      & {\small $2.243$}{\tiny ${}_{-0.053}^{+0.053}$} \\[0.8ex] 
{\small $\Omega_c h^2$}               & {\small $0.1114$}{\tiny${}_{-0.0060}^{+0.0061}$}      & {\small $0.1046$}{\tiny ${}_{-0.0049}^{+0.0047}$}      
                                      & {\small $0.1089$}{\tiny ${}_{-0.0041}^{+0.0041}$} \\[0.8ex]
{\small $\theta$}                     & {\small $1.039$}{\tiny${}_{-0.003}^{+0.003}$}         & {\small $1.037$}{\tiny ${}_{-0.003}^{+0.003}$}        
                                      & {\small $1.039 $}{\tiny ${}_{-0.003}^{+0.003}$} \\[0.8ex]
{\small $\tau$}                       & $\quad${\small$0.08712$}{\tiny${}_{-0.00721}^{+0.00565}$}$\quad$    & $\qquad${\small $0.08646$}{\tiny ${}_{-0.0067}^{+0.0061}$}$\quad$  
                                      & {\small$0.08797$}{\tiny${}_{-0.00627}^{+0.00618}$}  \\[0.8ex]
{\small $10^{8} \Sigma_I\,{}^b$}              & {\small $2.910$}{\tiny${}_{-1.229}^{+1.169}$}         & {\small $0.4845$}{\tiny ${}_{-0.3824}^{+0.2980}$}  
                                      & $--$  \\ [0.8ex] 
{\small $\Sigma_{II}\,{}^b$}                & {\small $-7.169$}{\tiny ${}_{-1.959}^{+2.218}$}       & $--$                             
                                      & $--$ \\[0.8ex]
{\small $n_s$}                        & {\small $0.9869$}{\tiny ${}_{-0.0184}^{+0.0192}$}     & {\small $0.9551$}{\tiny ${}_{-0.0137}^{+0.0135}$} 
                                      & {\small $0.9651$}{\tiny ${}_{-0.0124}^{+0.0123}$} \\[0.8ex]
{\small $\log[10^{10} A_s]$}$\quad$   & {\small $3.118$}{\tiny ${}_{-0.051}^{+0.051}$}        & {\small $3.039$}{\tiny ${}_{-0.040}^{+0.040}$} 
                                      & {\small $3.070$}{\tiny ${}_{-0.033}^{+0.031}$} \\[0.8ex]
{\small $A_{SZ}\,{}^c$}               & {\small $1.054 \pm 0.578$}                            & {\small $0.9544 \pm 0.5911$} 
                                      & {\small $1.040 \pm 0.574$} \\ [1ex]
\hline 
{\small $\Omega_d$}          & {\small $0.952$}{\tiny ${}_{-0.033}^{+0.033}$}              & {\small $0.956$}{\tiny ${}_{-0.030}^{+0.031}$} 
                             & {\small $0.955$}{\tiny ${}_{-0.027}^{+0.027}$}\\ [0.8ex] 
{\small $\mathcal{K}$}       & {\small $0.296$}{\tiny ${}_{-0.041}^{+0.042}$}              & {\small $0.270$}{\tiny ${}_{-0.036}^{+0.034}$}   
                             & {\small $0.291$}{\tiny ${}_{-0.032}^{+0.034}$} \\ [0.8ex]
{\small $t_0$}               & {\small $13.64$}{\tiny ${}_{-0.13}^{+0.12}\,$} {\small Gyr} & {\small $13.85$}{\tiny ${}_{-0.12}^{+0.11}\,$} {\small Gyr} 
                             & {\small $13.79 $}{\tiny ${}_{-0.11}^{+0.12}\,$}{\small Gyr} \\ [0.8ex]
{\small $\Omega_{\Lambda}$}  & {\small $0.734$}{\tiny ${}_{-0.024}^{+0.024}$}              & {\small $0.754$}{\tiny ${}_{-0.021}^{+0.022}$} 
                             & {\small $0.740$}{\tiny ${}_{-0.020}^{+0.019}$}\\[0.8ex]
{\small $H_0\,{}^d$}         & {\small $71.55$}{\tiny${}_{-1.91}^{+1.86}$}                 & {\small $71.95$}{\tiny ${}_{-1.96}^{+2.09}$} 
                             & {\small $71.14$}{\tiny ${}_{-1.85}^{+1.71}$} \\ [0.8ex]
\hline
\end{tabular}

{\small Notes.}

{\tiny

a. The mean values of the posterior distribution for each parameter. The quoted errors show the $68 \%$ confidence levels. 

b. $\Sigma_I$ and $\Sigma_{II}$ are given in units of the Thomson scattering cross section times the speed of light, 
   $\sigma_T = 6.65 \times 10^{-25}\,\text{cm}^2$ and $c=1$. 

c. The quoted errors in $A_{sz}$ are the standard deviations of the distributions.

d. $H_0$ is given in Km/s/Mpc.

}
\label{table:nonlin} 
\end{table*}

\begin{figure}
\begin{center}
\includegraphics[width=2.4in]{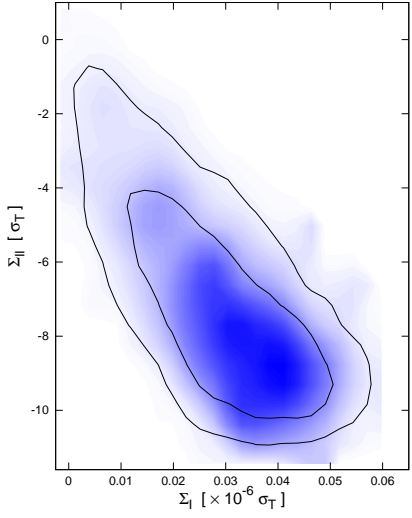}
\caption{Contour confidence intervals for $\Sigma_I$ vs $\Sigma_{II}$ at $68 \%$ and $95 \%$ c.l.  The shading shows the mean likelihood of the samples}
\end{center}
\label{fig4}
\end{figure}

\begin{figure}
\begin{center}
\includegraphics[width=1.6in]{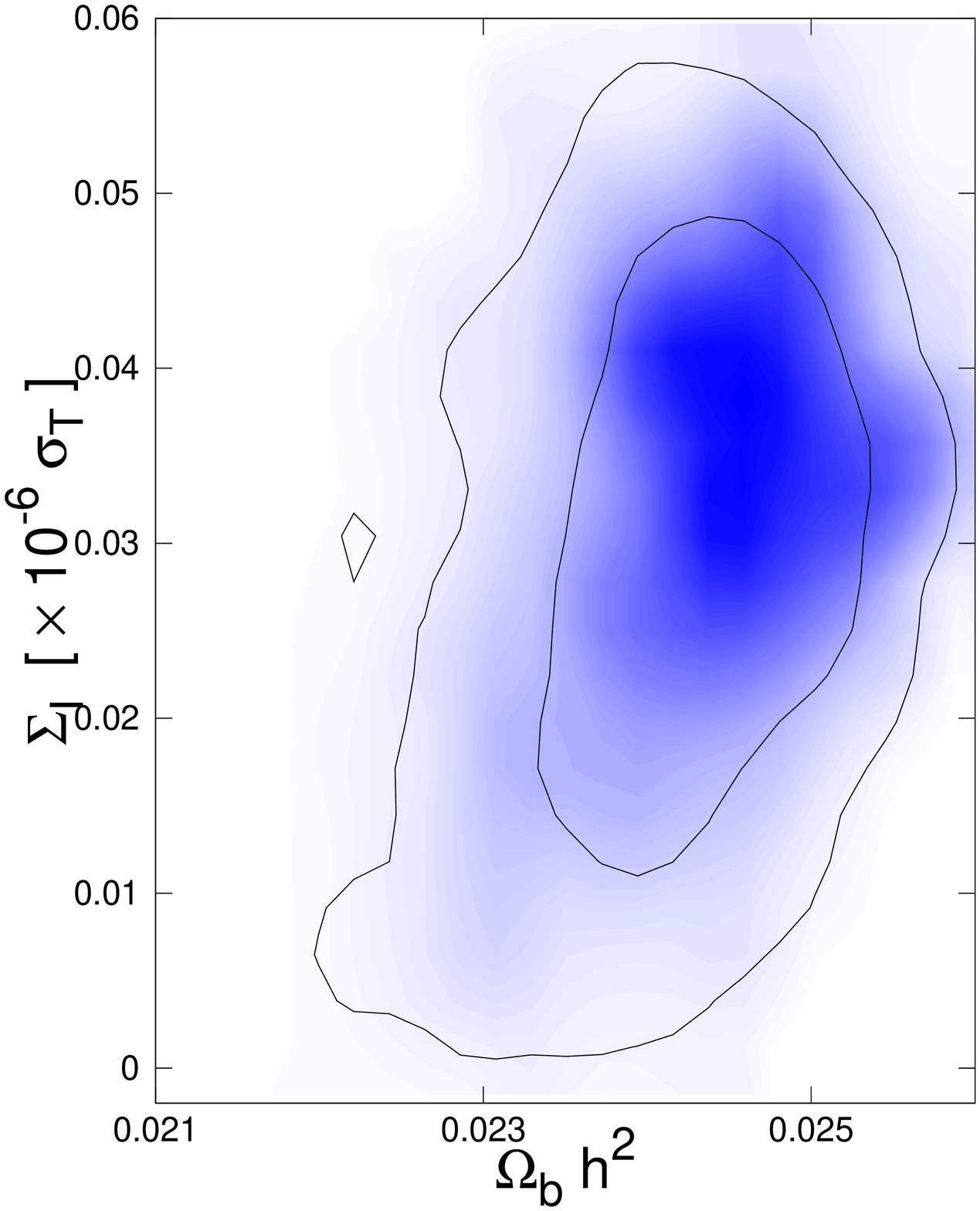}
\includegraphics[width=1.6in]{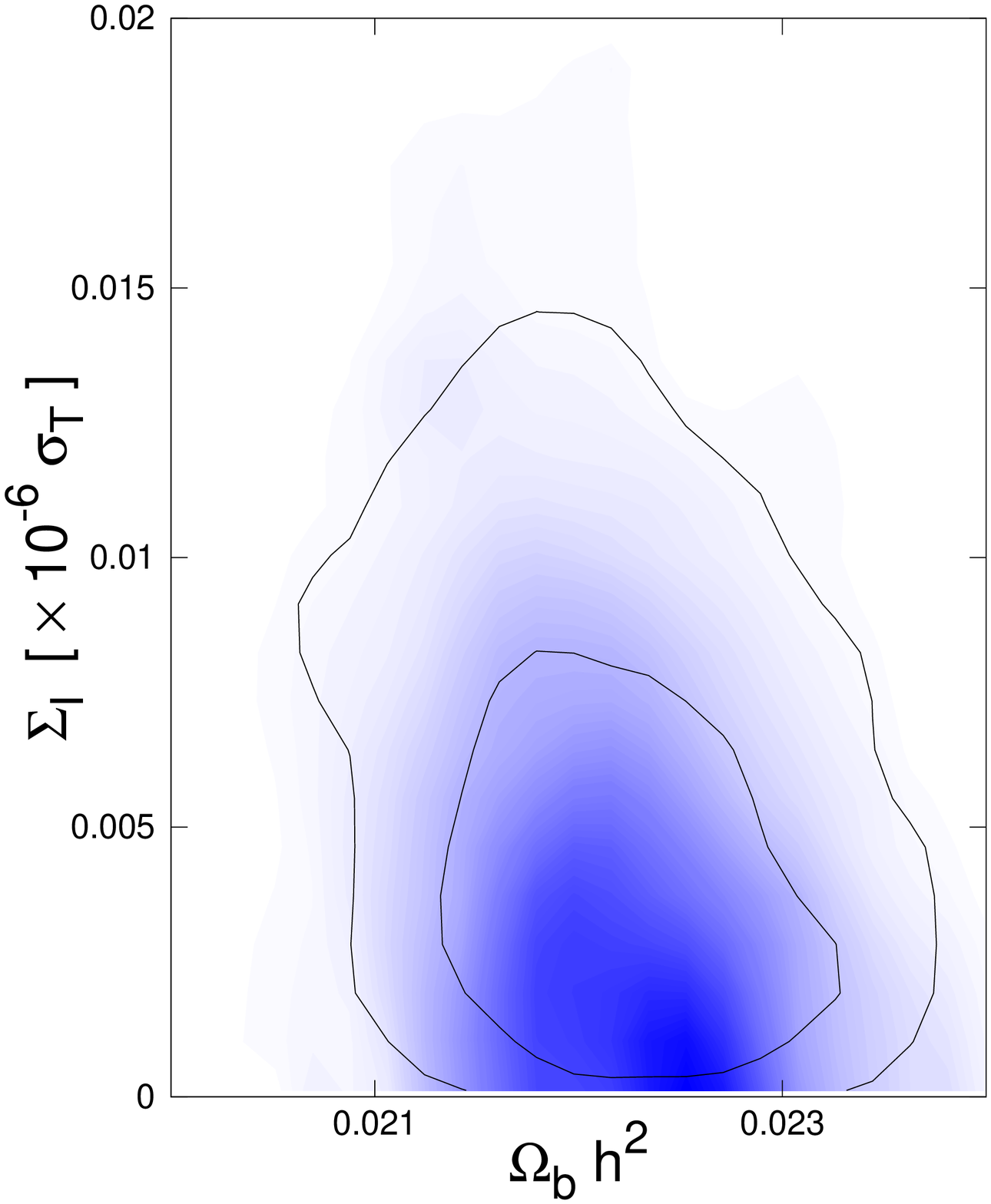} 
\caption{2D Posterior confidence intervals for  $\Omega_b h^2$ vs $\Sigma_I$ at $68 \%$ and $95 \%$ c.l. 
Left panel: Considering both interactions (model A). Right panel: considering only
interaction $\Sigma_I$ (model B). The shading shows the mean likelihood of the samples}
\end{center}
\label{fig5}
\end{figure} 

\begin{figure}
\begin{center}
\includegraphics[width=3in]{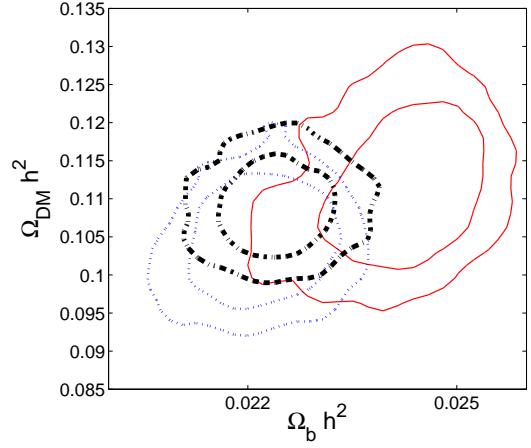}
\caption{Contour confidence intervals for $\Omega_b h^2$ vs $\Omega_{\text{DM}} h^2$ at $68 \%$ and $95 \%$ c.l. Solid (red) lines are for Model A, dot-dashed (blue)
for Model B, and thick dashed (black) for Model C.}
\end{center}
\label{fig6}
\end{figure}

\begin{figure}
\begin{center}
\includegraphics[width=3in]{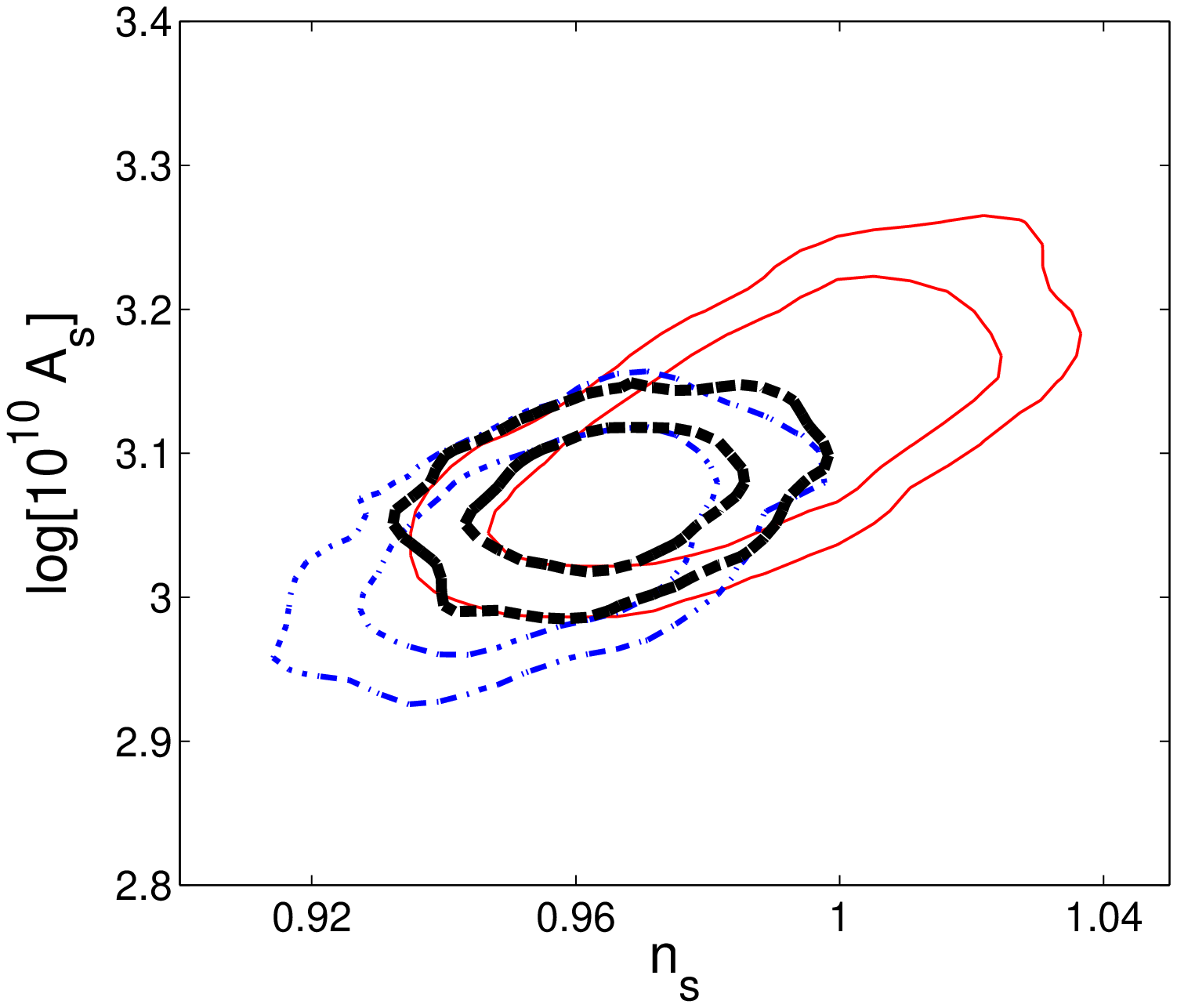}
\caption{Contour confidence intervals for $n_s$ vs $A_s$ at $68 \%$ and $95 \%$ c.l.  Solid (red) lines are for Model A, dot-dashed (blue)
for Model B, and thick dashed (black) for Model C.}
\end{center}
\label{fig7}
\end{figure}

In Fig. 2 we show the angular power spectrum for the 
CMB radiation. There several curves are shown for different values of $\Sigma_I$, which is expressed in units of $10^{-6}$ times the Thomson scattering 
cross section times the speed of light ($c=1$). 
We note that the greatest differences with the $\Lambda$CDM model comes at  high multipoles, which is not surprising because the interaction 
quickly decays and only modes which have entered into the horizon at early times in the Universe were  affected by it.

On the other hand, it is well known that interactions between the dark sector and baryons are tightly constrained from equivalence principle and  
solar system tests \cite{weak}. Nonetheless, such  interactions could be originated by  general predictions of string theory \cite{Damour90,Damour94}, leading to the necessity 
of some  screening mechanism \cite{khoury1_1,khoury1_2,khoury2,vainshtein} to evade these experimental constraints. 
In chameleon theories \cite{khoury1_1,khoury1_2}, the range of this interaction depends on the energy density of the surrounding medium, leading to long-range forces 
when the density is 
low, and to short-range forces when it is high. Inspired by these theories, we give  a $k$ dependence to  the hydrodynamical equations, such that the interactions 
respond to an effective wavenumber given by $k_{\text{eff}}^2 = k^2 g(\rho)$, where $g$ is a monotonic growing function of the ambient energy density.  
By noting that $\rho$ falls with the scale factor, we choose a potential law for the wavelength number,  $k_{\text{eff}}^2 = k^2/a^n$. In the following we specialize to the case $n=2$. 
We use the substitution $k \rightarrow k_{\text{eff}}$ in Eq. (\ref{interaction}) and obtain\footnote{Chameleon theories are modifications of gravity and their 
perturbation equations are different from those used here; it is out of the scope of this work to treat the precise equations of the chameleons. For such a treatment see
\cite{Khoury_Brax,Gannouji}.}

\begin{equation}
 f_{d} = \rho_d(1+w_{d}) \Sigma_{II}\frac{\rho_{d0}}{m_p} (\theta_{b} - \theta_d) / k^2.  \label{interaction2}
\end{equation}
Clearly, we can also understand this interaction as a dependence of the cross section on the ambient energy density. 
In Fig. 3 we show the CMB angular power spectrum for different values of the parameter $\Sigma_{II}$. We note that, in remarkable 
contrast to the $\Sigma_I$ interaction, here all  scales are  affected in a similar way. This is  because  the effective interaction remains constant 
while the scale factor grows and all modes feel it when they enter  the horizon.

\vspace{0.5cm}

\begin{figure}
\includegraphics[width=3.4in]{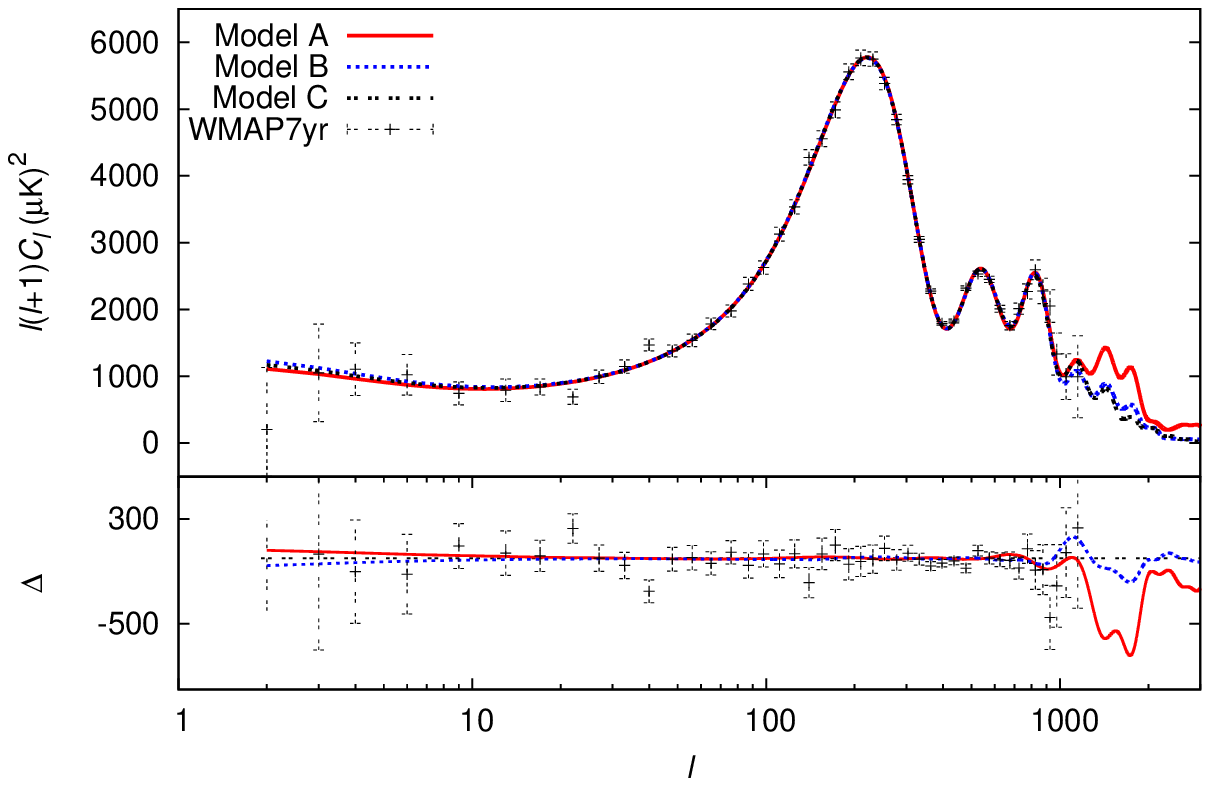}
\caption{CMB power spectrum for the mean values estimated with the MCMC analysis by CosmoMC for the three models. The reported difference is $\Delta = 
l(l+1)\left(C_l^{\text{Model C}} - C_l \right) $.}
\label{fig9}
\end{figure}

It is possible to treat both parametrizations together if we choose

\begin{equation}
 f_{d} = \rho_d(1+w_{d})  \frac{(\Sigma_{I} + \Sigma_{II}a^2)\rho_{d0}}{m_p a^2}   (\theta_{b} - \theta_d) / k^2.  \label{interaction3}
\end{equation}
The hydrodynamical equations become

\begin{eqnarray}
\dot{\delta}_d &=& - (1+ w_d)(\theta_d - 3 \dot{\Phi}) + 3 \mathcal{H}  w_d \delta_d,
                              \label{pertddfc2} \\
\dot{\theta}_d &=& - \mathcal{H} \theta_d  + k^2 \Psi -  \frac{(\Sigma_{I} + \Sigma_{II}a^2)\rho_{d0}}{m_p a^2}  (\theta_{b} - \theta_d),  \label{perttdfc2}
\end{eqnarray}
and for baryons
\begin{eqnarray}
 \dot{\delta}_b &=& -\theta_b + 3 \dot{\Phi},  \label{pertdbc2} \\
 \dot{\theta}_b &=& - \mathcal{H} \theta_b  + k^2 \Psi + c^2_{sb} k^2 \delta_b \nonumber\\ 
                & &  - \frac{\rho_d}{\rho_b} (1+w_{d}) \frac{(\Sigma_{I} + \Sigma_{II}a^2)\rho_{d0}}{m_p a^2} 
                       (\theta_{d} - \theta_b).  \label{perttbc2}
\end{eqnarray}
If one assumes the $\Lambda$CDM  decomposition of the dark fluid in dark matter and cosmological constant, it is easy to see that instead of Eqs. (\ref{pertddfc})
and (\ref{perttdfc}), the following  equations govern the evolution of the dark matter perturbation

\begin{eqnarray}
 \dot{\delta}_{\text{DM}} &=& -\theta_{\text{DM}} + 3 \dot{\Phi} + \frac{\delta q_{\text{DM}}}{\rho_{\text{DM}}},  \label{pertddmc2} \\
 \dot{\theta}_{\text{DM}} &=& - \mathcal{H} \theta_{\text{DM}}  + k^2 \Psi  + \frac{k^2 f_{\text{DM}}}{\rho_{\text{DM}}}, \label{perttdmc3}
\end{eqnarray}
where we have used the conditions (\ref{ass1}) and (\ref{ass2}). The transfer energy and momentum  terms are related by

\begin{equation}
 \delta q_{\text{DM}}= \delta q_{d}   \qquad \text{and} \qquad  f_{\text{DM}} = f_{d}.
\end{equation}
Thus, although
the degeneracy with the $\Lambda$CDM has been broken at first order in perturbation theory, there exist degeneracies to other models, as in this case to 
$\Lambda$CDM with the same interactions, which means that the general class of interactions given by Eqs. (\ref{pertddfc})
and (\ref{perttdfc}) does not help us to elucidate the actual decomposition of the dark fluid (if it exists).  Accordingly we can treat the above 
described couplings as interactions between dark matter and baryonic matter, without loss of generality, as we will
do for numerical purposes\footnote{Special care must be taken because numerical codes, as CAMB, use synchronous gauge and fix the residual gauge freedom by
taking $\theta_{\text{DM}} = 0$. In cases in which there are sources in the $\theta$ evolution equation, like ours, this is not possible to do.}. 
Thereafter, if desired, we can go back and forth between both models'  results using Eqs. (\ref{deg1rel}) and (\ref{deg2rel}).

To constrain the interactions, we perform a Monte Carlo Markov Chain (MCMC) analysis over the nine-parameter space (Model A) 
$\{ \Omega_b h^2, \Omega_{\text{DM}} h^2, \theta, \tau, n_s, \log A_s, A_{sz}, \Sigma_I, \Sigma_{II} \}$ using the code CosmoMC \cite{Lewis2002}. 
$\theta$ is the ratio of the sound horizon to the angular diameter distance at recombination,
$\tau$ is the reionization optical depth,  $n_s$ is the spectral index of the primordial scalar perturbations and $A_s$ is its amplitude at a pivot scale of 
$k_0 = 0.05\, \text{Mpc}^{-1}$. 

We have imposed flat priors on the two interaction parameters: $0<\Sigma_I< 10^{-7 } \times \sigma_T$ and $-11 \times \sigma_T <\Sigma_{II}< 10 \times \sigma_T$. 
For the CMB anisotropies and polarization data
we used the Wilkinson Microwave Anisotropy Probe (WMAP) seven-year observations results \cite{Larson}. For the joint analysis we use also Hubble  
Space Telescope measurements (HST) \cite{Riess2009} to impose a Gaussian prior on the Hubble constant today of $H_0 = 74 \pm 3.6 \,\text{km/s/Mpc}$, 
and the supernovae type Ia Union 2 data set compilation by the Supernovae Cosmology Project \cite{Union2}.

We also study two other models: Model B, only considering the interaction $\Sigma_I$, it has an eight-parameter space  
$\{ \Omega_b h^2, \Omega_{\text{DM}} h^2, \theta, \tau, n_s, \log A_s, A_{sz}, \Sigma_I \}$; 
and Model C, which does not consider any interaction, a seven-parameter space  
$\{ \Omega_b h^2, \Omega_{\text{DM}} h^2, \theta, \tau, n_s, \log A_s, A_{sz} \}$, corresponding to the standard $\Lambda$CDM model.

The summary of the posterior one-dimensional marginalized probabilities is outlined in Fig. 4 and Table \ref{table:nonlin}. In Fig. 5 we show the contour 
confidence intervals for the  marginalized $\Sigma_I\!-\!\Sigma_{II}$ space at 0.68 and 0.95 confidence levels (c.l.). There,  the
high degeneracy between both parameters is shown: while $\Sigma_{II}$ takes values closer to zero, $\Sigma_I$ also does.
It is interesting that nonzero values of the interactions (when introduced) are consistent and preferred by the considered data at 0.95 c.l.

We note that the addition of the $\Sigma_I$ and $\Sigma_{II}$ interactions to the theory produce remarkable differences between the parameters 
estimations of Model A and Model C. 
This is more evident for the baryonic matter energy density $\Omega_b h^2$, as can be observed from Figs. 4, 6,  and 7, or read directly from Table I.
For this reason, we study Model B, which only considers the $\Sigma_I$ interaction. In this case, the tensions between the parameters estimations are alleviated.
The discrepancies are also evident in the primordial power spectrum parameters $A_s$ and $n_s$, which also can be seen from Figs. 4 and 8. Nonetheless, these are not 
alleviated when one considers Model B.

Instead of using the proton mass as the scale in the interactions, we can use an  arbitrary associated mass to the dark matter or dark fluid  particles, $m_d$. 
We obtain the following constraints at 0.68 c.l. on the ratio $\Sigma/m_d$ (we use $c = 3 \times 10^{10} \text{cm/s}$):

For the case in which we consider both interactions (Model A)
\begin{equation}
3.58 \times 10^{-22} <  \frac{\Sigma_I}{m_d} < 8.68 \times 10^{-22} \,\,\frac{\text{cm}^3/\text{s}}{\text{GeV}/c^2},
\end{equation}
and
\begin{equation}
-1.94  \times 10^{-13}  <  \frac{\Sigma_{II}}{m_d} <  -1.05 \times 10^{-13} \,\, \frac{\text{cm}^3/\text{s}}{\text{GeV}/c^2}.
\end{equation}
While for Model B,
\begin{equation}
0.22 \times 10^{-22} <  \frac{\Sigma_I}{m_d}<  1.66 \times 10^{-22} \,\, \frac{\text{cm}^3/\text{s}}{\text{GeV}/c^2}. 
\end{equation}

We only obtained bounds of $m_d$ and $\Sigma$ in the combination $\Sigma/m_d$, although it is possible to find constrictions of them separately, even to the cross
section and the particles' thermal velocity, leading to constraints in the $\sigma \!-\! m_d$ plane. Nevertheless, to do this we have to allow
$c_s^2 \neq 0$, and moreover, make further assumptions on the thermodynamics of the dark matter (or the dark fluid)  \cite{Chen2002}. 

Note that the {\it effective} thermalized cross section for interaction $II$ is $a^2 \Sigma_{II}$, and this is equal  to  $\Sigma_I$ at about a redshift $ z \sim 10^5$.
Before this epoch interaction $I$ dominates, and after that, interaction $II$ starts to do it.  Just before recombination, at $z \simeq 1100$, interactions $I$ and $II$ are  
smaller  in strength than the Thomson interaction by about 9 and 5 orders of magnitude, respectively. After this time, the ionization fraction $x_e$ falls exponentially and
Thomson scattering becomes subdominant very quickly.

In Fig. 9 we plot the CMB power spectrum for the mean values estimated by the MCMC analysis. The reported difference is $\Delta = 
l(l+1)\left(C_l^{\text{Model C}} - C_l \right) $. We note that the three models are well inside the error bars of the binned measurements 
from WMAP seven-year observations results. Nonetheless, at higher multipoles ($l>1000$) the three models start to have notable discrepancies. The higher differences appear in
the region $1000<l<2000$, just inside the window  in which the primordial power spectrum parameters are expected to be estimated with higher precision ($1000<l<3000$). 
For $l>3000$ secondary anisotropies, mainly the Sunyaev-Zeldovich effect, are expected to dominate the CMB spectrum, 
while for multipoles $l<1000$ the spectrum is more sensible to the other parameters.  
These high-$l$ power spectrum multipoles will be 
probed by the PLANCK mission \cite{PLANCK2011}, so we expect to obtain tighter constraints in the near future.   

In this work we have not considered the effect that interactions $I$ and $II$ could have on big bang nucleosynthesis. 
This is because our phenomenological model only includes the
thermalized cross sections $\Sigma_I$ and $\Sigma_{II}$ of elastic collisions, whose strengths are at least 9 orders of magnitude weaker than the Thomson scattering at this epoch, 
and more important, do not annihilate baryons and maintain the baryon-to-photon ratio unaltered.  Accordingly,  we expect the effect over this process to be quite weak.

\section{Conclusions}

In this paper we worked out some properties of the so-called dark degeneracy \cite{Kunz09}: the fact that we can only measure the total energy momentum tensor 
of the dark sector and any split into different pieces (as in dark matter and dark energy) is artificial. Although it could be mathematically convenient. 
 
We start by defining the {\it dark fluid} as a barotropic fluid with speed of sound equal to zero, as in \cite{Beca2007,Balbi07} and, more recently, in 
\cite{Quevedo} (for similar approaches see \cite{Linder09,Arbey2005}). This is motivated by astrophysical constraints in the
dark matter besides its  cosmological properties. 
Making the speed of sound equal to zero allows dark fluid energy density perturbations to grow at all length scales, as a cold dark matter component does,
while it leaves room for a constant pressure different from zero. Astrophysical scenarios forbid this pressure to be very high, but it well could take values of
the cosmological critical density today. From the positivity of the  energy density of the dark fluid it follows that this pressure has 
to be negative. Thus, we conclude that the dark fluid could  act as  dark energy. Then we study the background cosmology for the dark fluid and show that, not surprisingly, 
the same evolution equations as in the $\Lambda$CDM model are obtained, leading to a degeneracy between both models. 
We conclude that any collection of fluids for which its total equation of state parameter is equal to Eq. (\ref{EOST2}) will lead us to the same result at the 
background cosmological level.

This is not necessarily the case when we consider linear perturbations. In order 
to preserve the dark degeneracy  we demand that the dark fluid be a perfect fluid. For a barotropic fluid, this last condition implies 
that the first order space-space part of the energy 
momentum tensor is equal to zero. Furthermore, one has to add that the energy momentum flux density is equal to that of the $\Lambda$CDM. Thus, we have
demanded that the energy momentum tensor of the dark fluid be equal to the $\Lambda$CDM one  at first-order in perturbation theory, otherwise the degeneracy is broken.

When considering a collection of multiple interacting fluids and demanding that the adiabatic speed of sound of the total composed fluid be 
equal to zero, we obtain exactly the perturbation equations of the dark fluid. This shows that the degeneracy is present also for more complicated
dark sector schemes. 

From  the first five sections, we conclude that there exist an infinity of cosmological models that give exactly the same 
observational signatures. Accordingly, it is fundamentally impossible to elucidate the actual structure of the dark sector. Thus, it is 
a matter of taste to prefer the $\Lambda$CDM model over any one of the exposed in this paper. In fact, to economize it is better to take the one 
fluid choice (the dark fluid) as the correct model. 

The only hope we have to understand the actual nature of the dark sector, is that it interacts in some way with the particles of the standard model. 
Of course interactions  would in general break the degeneracy, but this is not necessarily true. This subject is studied in Sec. VI, where we allow the  
degeneracy of the dark fluid and the $\Lambda$CDM model
in the homogeneous and isotropic cosmology, and then, we try to break it by adding interactions to baryons at first order in perturbations theory. We show that 
a general class of interactions defined by Eqs. (\ref{pertddfc}) and (\ref{perttdfc}) also can be understood as  between dark matter and baryons in 
the $\Lambda$CDM model context, leading to a new degeneracy, 
although not with the concordance model, but with a  $\Lambda$CDM-plus-interactions model. As a consequence, these interactions does not help us to favor any decomposition
of the dark fluid.

For the latter investigation we have chosen two independent interactions: one that resembles the Thomson scattering between photons and baryons, regulated by a 
parameter $\Sigma_I$, and a second, density-dependent interaction, inspired by chameleon theories and parametrized by $\Sigma_{II}$. For the analysis
we have used a combination of the seven-year results of WMAP, the Union 2 data set compilation of supernovae measurements, and a prior in the Hubble 
constant of the Hubble Space Telescope. 

When we consider both interactions together, we obtain that other parameters of the theory differ too much from its standard values, up to almost  $10 \%$ 
in the case of the baryonic density parameter $\Omega_b h^2$. Then, we study the case in which $\Sigma_{II}$ is equal to zero, allowing $\Sigma_I$ to vary 
(Model B). Here,  we show that the tension between the parameters estimation is reduced.

The introduction of the two parameterized interactions is allowed by current data, and if done, it is remarkable that nonzero values of them are preferred
at 0.95 confidence level.

\begin{acknowledgments}
JLCC  acknowledges  CONACYT for grant No. 84133-F.
AA acknowledges  CONACYT for grant no. 215819.
\end{acknowledgments}

\end{document}